\documentclass[twocolumn,amsmath,amssymb,nofootinbib,a4paper]{revtex4} 

\usepackage[utf8]{inputenc}
\usepackage[T1]{fontenc}
\usepackage{newunicodechar}
\usepackage{graphicx} 
\usepackage{amsmath}
\usepackage{amsfonts,amsbsy}
\usepackage{amssymb}
\usepackage{subdepth}
\usepackage{bbold}
\usepackage{nicefrac}
\usepackage[dvipsnames]{xcolor}
\usepackage{mathrsfs}
\definecolor{lcolor}{rgb}{0.5,0,0}
\definecolor{citcolor}{rgb}{0,0.3,0.0}
\usepackage[breaklinks,colorlinks,urlcolor=blue,citecolor=citcolor,linkcolor=lcolor]{hyperref}
\usepackage{braket, bbold, empheq, pdflscape, comment, pbox, leftidx}
\usepackage[normalem]{ulem}

\graphicspath{{Images/}}

\newcommand{\nc}{{N_\mathrm{c}}}

\newcommand{\as}{\alpha_\mathrm{s}}

\newcommand{\nr}[1]{(\ref{#1})} 
\newcommand{\fig}{Fig.~}

\newcommand{\eq}{Eq.~}

\newcommand{\eqs}{Eqs.~}

\DeclareMathAlphabet{\ib}{OML}{cmm}{b}{it}

\newcommand{\uu}[2]{U_{\ib{{#1}}{{#2}}}}
\newcommand{\uub}[2]{\bar{U}_{\bar{\ib{{#1}}}{{#2}}}}
\newcommand{\ud}[2]{U^\dagger_{\ib{{#1}}{{#2}}}}

\newcommand{\ua}[3]{\tilde{U}^{{#1}}_{\ib{{#2}}{{#3}}}}
\newcommand{\uab}[3]{\bar{\tilde{U}}^{{#1}}_{\bar{\ib{{#2}}}{{#3}}}}
\newcommand{\tr}[1]{\mathrm{tr}\left\{{{#1}}\right\}}
\newcommand{\epr}[2]{e^{ i \epsilon g \alpha^R_{\ib{{{#1}}},{{#2}}}}}
\newcommand{\emr}[2]{e^{ - i \epsilon g \alpha^R_{\ib{{{#1}}},{{#2}}}}}
\newcommand{\epl}[2]{e^{ i \epsilon g \alpha^L_{\ib{{{#1}}},{{#2}}}}}
\newcommand{\eml}[2]{e^{ - i \epsilon g \alpha^L_{\ib{{{#1}}},{{#2}}}}}
\newcommand{\ar}[2]{\alpha^R_{\ib{{{#1}}},{{#2}}}}
\newcommand{\al}[2]{\alpha^L_{\ib{{{#1}}},{{#2}}}}
\newcommand{\la}[3]{\lambda_{\ib{{#1}}{{#2}}}^{{#3}}}
\newcommand{\lab}[3]{\bar{\lambda}_{\bar{\ib{{#1}}}{{#2}}}^{{#3}}}

\newcommand{\ld}[2]{L^{{#1}}_{\ib{{{#2}}}}}
\newcommand{\rd}[2]{R^{{#1}}_{\ib{{{#2}}}}}
\newcommand{\ldb}[2]{\bar{L}^{{#1}}_{\bar{\ib{{{#2}}}}}}
\newcommand{\rdb}[2]{\bar{R}^{{#1}}_{\bar{\ib{{{#2}}}}}}
\newcommand{\ldalt}[3]{L^{{#1}}_{\ib{{{#2}}},#3}}
\newcommand{\rdalt}[3]{R^{{#1}}_{\ib{{{#2}}},#3}}
\newcommand{\ldbalt}[3]{\bar{L}^{{#1}}_{\bar{\ib{{{#2}}}},#3}}
\newcommand{\rdbalt}[3]{\bar{R}^{{#1}}_{\bar{\ib{{{#2}}}},#3}}

\newcommand{\diag}[2][{}]{\pbox{\textwidth}{\includegraphics[scale=0.5]{{#2}}}}

\begin{document}

\title{Unequal rapidity correlators in the dilute limit of JIMWLK}

\author{T. Lappi}
\email{tuomas.v.v.lappi@jyu.fi}
\affiliation{
Department of Physics, %
 P.O. Box 35, 40014 University of Jyv\"askyl\"a, Finland
}
\affiliation{
Helsinki Institute of Physics, P.O. Box 64, 00014 University of Helsinki,
Finland
}

\author{A. Ramnath}
\email{andrecia.a.ramnath@student.jyu.fi}
\affiliation{
Department of Physics, %
 P.O. Box 35, 40014 University of Jyv\"askyl\"a, Finland
}

\begin{abstract}
We study unequal rapidity correlators in the stochastic Langevin picture of Jalilian-Marian--Iancu--McLerran--Weigert--Leonidov--Kovner (JIMWLK) evolution in the Color Glass Condensate effective field theory. We discuss a diagrammatic interpretation of the long-range correlators. By separately evolving the Wilson lines in the direct and complex conjugate amplitudes, we use the formalism to study two-particle production at large rapidity separations. We show that the evolution between the rapidities of the two produced particles can be expressed as a linear equation, even in the full nonlinear limit. We also show how the Langevin formalism for two-particle correlations reduces to a BFKL picture in the dilute limit and in momentum space, providing an interpretation of BFKL evolution as a stochastic process for color charges.
\end{abstract}

\maketitle

\section{Introduction}

Multi-particle correlations, both in azimuthal angle and rapidity, are becoming an increasingly important experimental tool to access the properties of QCD in high energy collision systems. There is an intensive debate (see e.g. \cite{Schlichting:2016sqo}) on the origin of the structure of azimuthal correlations in small (proton-proton and proton-nucleus) collision systems, where QCD correlations already present in the colliding objects compete with the effects of particle reinteractions (such as hydrodynamical flow or escape bias) in the later stage of the collision. In order to fully sort out the interpretation of the experimental results, it is important to fully understand the QCD dynamics leading to particle correlations. In particular, a characteristic feature of a hydrodynamical-like azimuthal correlation in a hadronic collision is that it extends far in rapidity~\cite{Dumitru:2008wn}. Whereas effects at later times, such as resonance decays, have a very short range in rapidity, effects that extend to large rapidity separations must originate early in the collision. This is the case both for hydrodynamical flow, which is sensitive to the coordinate space geometry that is similar at all rapidities, and for correlations between the partons in the colliding hadrons. Therefore, one needs to understand in QCD how the production of particles in high energy collisions is correlated across large rapidity separations. 

The Color Glass Condensate (CGC, see e.g. \cite{Iancu:2003xm,Weigert:2005us,Gelis:2010nm}) is an effective theory of QCD for high energy processes. It is based on a separation of scales between ``fast'' large-$x$ degrees of freedom that are integrated out into an effective description and the soft small-$x$ gluons that are the relevant degrees of freedom for high energy scattering. The longitudinal momentum cutoff separating these two scales can be varied, and the effect of changing it absorbed into a renormalization of the effective description as a function of the longitudinal momentum (or rapidity) scale. This procedure leads to the JIMWLK\footnote{
	The acronym stands for Jalilian-Marian--Iancu--McLerran--Weigert--Leonidov--Kovner.} evolution equation~\cite{Jalilian-Marian:1997xn,Jalilian-Marian:1997jx,Jalilian-Marian:1997gr,Jalilian-Marian:1997dw,JalilianMarian:1998cb,Weigert:2000gi,Iancu:2000hn,Iancu:2001md,Ferreiro:2001qy,Iancu:2001ad,Mueller:2001uk}, 
which can be used to resum leading logarithmic (in energy or $x$) corrections to QCD scattering cross sections. In practical calculations, the convenient degree of freedom for describing high energy QCD scattering is the Wilson line (see e.g. \cite{Buchmuller:1995mr}). This is the eikonal scattering amplitude for a partonic probe passing through the target color field. In the CGC picture, the Wilson lines at each point in the transverse plane are stochastic variables drawn from a probability distribution; it is this probability distribution whose dependence on the rapidity $Y= \ln 1/x$ is given by the JIMWLK equation. An equivalent formulation for the evolution of the probability distribution is provided by the Langevin formulation~\cite{Blaizot:2002xy}, where the Wilson lines themselves depend on rapidity through a stochastic partial differential equation involving a random noise. In addition to providing a more direct physical picture of the evolution, the Langevin formulation is the basis for numerical solutions of the JIMWLK equation~\cite{Rummukainen:2003ns,Kovchegov:2008mk,Lappi:2012vw}.

The most common phenomenological applications of the CGC framework involve processes in which one needs only the Wilson lines at one rapidity. This includes DIS cross sections~\cite{Bjorken:1970ah,Nikolaev:1990ja,Nikolaev:1991et,Beuf:2016wdz,Beuf:2017bpd,Hanninen:2017ddy}, where the relevant rapidity depends only on the energy of the incident virtual photon, and single inclusive particle production~\cite{Dumitru:2001jn,Dumitru:2002qt,Dumitru:2005gt,Chirilli:2011km,Chirilli:2012jd,Kang:2014lha,Ducloue:2016shw,Iancu:2016vyg,Ducloue:2017mpb,Ducloue:2017dit}, where it is determined by the kinematics of the produced particle. This is also true for multi-particle production~\cite{Marquet:2007vb,Lappi:2012nh}, if the produced particles are close to each other in rapidity.
The situation becomes more complicated, however, if one is interested in the correlations between particles that are separated by a parametrically large rapidity interval $\Delta Y \gtrsim 1/\as$.

For the case of two dense projectiles, there is a formalism, developed initially in \cite{Gelis:2008sz}, in which one follows a separate JIMWLK evolution for each of the colliding nuclei (see also \cite{Lappi:2009fq} for a discussion). While the theoretical status of this formulation is still poorly understood (see e.g. the calculation reported in \cite{Lappi:2012gg} suggesting that the decorrelation speed in rapidity is not an infrared safe quantity in this framework), it has been used in some phenomenological applications~\cite{Schenke:2016ksl}. It has also led, via a $k_T$-factorized approximation, to the ``glasma graph'' or related calculations of multi-particle correlations in small collision systems~\cite{Dusling:2009ni,Kovner:2010xk,Kovner:2011pe,Dumitru:2010iy,Dusling:2012iga,Dusling:2012cg,Dusling:2012wy}. In a perturbative language, the correlation in these calculations originates from two particles being produced from different BFKL ladders. This contribution dominates when both the projectile and target are parametrically dense, so that there is no suppression for having additional ladders between them. In this sense, these are ``dense-dense'' calculations, even if done in a $k_T$-factorized approximation. On the other hand, when both the projectile and target are dilute, correlated semihard particle production should be dominated by production from a single ladder, or the ``jet graphs'' in the language of \cite{Dusling:2012cg,Dusling:2012wy}. Here, perturbative calculations of azimuthal decorrelations between these (Mueller-Navelet) jets have been performed, even at the NLO level~\cite{Ducloue:2013hia}. 

For correlations between particles with large rapidity separations, it turns out that the ``dilute-dense'' case is actually in some sense the most complicated one. In the power counting of \cite{Gelis:2008ad,Gelis:2008sz}, where the color charge density in the dense target is parametrically large $\rho \sim 1/g$, and in the dilute projectile parametrically small $\rho \sim g$, both the ``single ladder'' (or ``jet graph'') and ``separate ladder'' (or ``glasma graph'') contributions are parametrically (in $\as$) equally important. Out of these two kinds of contributions, there has been a lot of recent work in understanding the ``separate ladder'' contributions beyond the glasma graph approximation~\cite{Kovner:2010xk,Kovner:2011pe,Lappi:2015vta,Dusling:2017dqg,Mace:2018yvl}.

In the case of a dense target, our understanding of how to calculate the ``single ladder'' contribution is much less developed. 
One needs to generalize the fully nonlinear JIMWLK equation to the evolution of not just operators made out of Wilson lines at a single rapidity, but correlations of Wilson lines at different rapidities. For this purpose, a formalism based on the Langevin description of JIMWLK evolution was developed by Iancu and Triantafyllopoulos (IT) in \cite{Iancu:2013uva} (see also earlier, very similar work in \cite{Kovner:2006ge,Kovner:2006wr}). Here, one derives a new Langevin equation for a bilocal quantity that encodes the correlation between Wilson lines at two different rapidities. This formalism has not, however, been fully applied to phenomenology, nor has it been analyzed in more detail. 

Our intention in this paper is to do the latter, with the main purpose of elucidating the diagrammatic interpretation of the IT formalism. We do this by starting from the bilocal Langevin description and taking the dilute limit. We show explicitly how this procedure recovers a two-particle correlation originating in particle production from the same BFKL ladder. In the process, we show also that the bilocal Langevin formulation, even in the full nonlinear case, can actually be transformed into a form in which the evolution between the rapidities of the two produced particles is linear. This somewhat surprising, or even counter-intuitive, result seems to confirm what has been found earlier in \cite{JalilianMarian:2004da,Kovner:2006wr}. This statement does not mean that the two-particle production process would somehow be fully linear; one still needs to solve the nonlinear evolution equation for the Wilson lines themselves.  Rather than explore the full phenomenological consequences of this picture, we will try to elucidate the physics in this formalism and set the stage for such a calculation in future work. For concreteness, we will focus throughout this paper on the two-particle cross section specifically for the production of a quark and a gluon. The incoming projectile is consequently a Wilson line in the fundamental representation; the generalization of this to a gluon probe should be relatively straightforward.

This paper is structured as follows. First, we review the basics of JIMWLK evolution in Sec.~\ref{sec:jimwlk}, both in the Fokker-Planck and in the Langevin formulations. We then discuss in Sec. \ref{sec:equal} two-particle production at parametrically similar rapidities, i.e. without evolution between the rapidities of the particles. In Sec.~\ref{sec:dilute}, we develop the dilute limit of JIMWLK evolution in terms of color charges or, more accurately, Reggeized gluons, leading to the BFKL equation for the unintegrated gluon distribution. We then move in Sec.~\ref{sec:neqy} to the IT formalism of Langevin evolution between the rapidities of the two produced particles, slightly rewriting the evolution equation to highlight the linear structure in the evolution between the two rapidities. Finally, in Sec.~\ref{sec:dilute2part} we show how the dilute limit of the IT formalism leads to a factorized structure that one would expect from BFKL dynamics, with a BFKL Green's function separating the two particles.

\section{JIMWLK evolution}
\label{sec:jimwlk}

\subsection{The JIMWLK equation in the Fokker-Planck formalism}

We consider a high energy interaction of a dilute colored probe with the color field of a dense target. 
In the CGC theory, the expectation value of an observable $\hat{\mathcal{O}}$ that is local in rapidity $Y$, is given by
\begin{align}
\left\langle \hat{\mathcal{O}} \right\rangle_Y \equiv \int [DU] W_Y[U] \hat{\mathcal{O}}.
\label{exp-val}
\end{align}
Here $[DU]$ is the functional de Haar measure on SU$(\nc)$ and $W_Y[U]$ is the CGC weight function describing the density distribution at $Y$ of the Wilson lines in the target.
These Wilson lines $U = U(\ib{x}) \equiv \uu{x}{}$ are unitary, path-ordered exponentials
\begin{align}
\ud{x}{} \equiv P \exp \left\{ i g \int dx^+ \alpha^a_\ib{x}(x^+) t^a \right\},
\end{align}
represented diagrammatically following the notation of \cite{Kovchegov:2008mk,Marquet:2010cf,Lappi:2016gqe} as
\begin{align}
\pbox{\textwidth}{\includegraphics[scale=0.15]{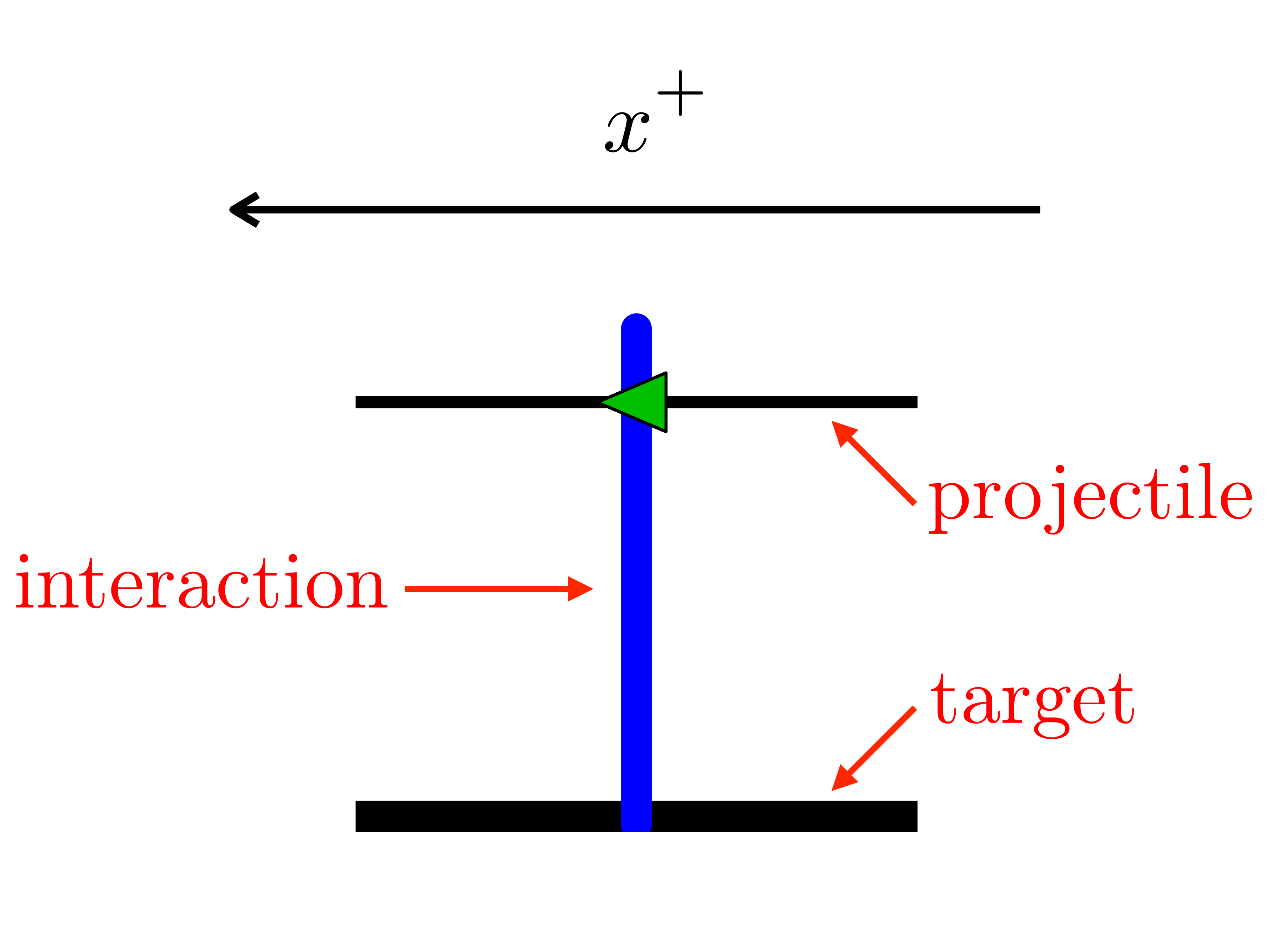}}.
\end{align}
Here, $\alpha^a_\ib{x}$ is the color field generated by the target with color index $a=1,...,\nc$ and the $t$'s are the fundamental generators of SU($\nc$). 
The lightcone time axis $x^+$ runs from right to left in these diagrams. The Hermitian conjugate Wilson line is then denoted by an arrow facing the opposite way:
\begin{align}
\uu{x}{} = \diag{empty} \diag{targref} \diag{empty}.
\end{align}
An example of a simple observable that is relevant in this context is the quark dipole
\begin{align}
\hat{S}_\ib{xy} \equiv \frac{\tr{\ud{x}{} \uu{y}{}}}{\nc}
= \frac{1}{\nc} \tr{\diag{empty2} \diag{targ2} \diag{empty2_xy}}.
\end{align}

The dependence of the target color field on rapidity is described by JIMWLK evolution.  Here the CGC weight function evolves from an initial condition $Y_\mathrm{in}$ to a final $Y$ according to the JIMWLK equation 
\begin{align}
\frac{\partial}{\partial Y} W_Y [U] = H W_Y[U].
\label{differential-eq}
\end{align}
Typically, a Gaussian distribution is used for the initial condition $W_{Y_\mathrm{in}}$, as in the McLerran-Venugopalan (MV) \cite{McLerran:1994ni,McLerran:1994ka,McLerran:1994vd} model. 
The JIMWLK Hamiltonian is
\begin{align}
H \equiv \frac{1}{8 \pi^3} \int _\ib{uvz} \mathcal{K}_\ib{uvz} (\ld{a}{u} - \ua{\dagger ab}{z}{} \rd{b}{u}) (\ld{a}{v} - \ua{\dagger ac}{z}{} \rd{c}{v}),
\label{ham}
\end{align}
where tildes denote the adjoint representation and two-dimensional coordinate space integrals are denoted with the shorthand $\int_\ib{u} \equiv \int d^2u$.
The JIMWLK kernel is 
\begin{equation}
\mathcal{K}_\ib{uvz} \equiv \mathcal{K}^i_\ib{uz} \mathcal{K}^i_\ib{vz}, 
\end{equation}
where 
\begin{equation}
\mathcal{K}^i_\ib{uz} = \frac{(\ib{u} - \ib{z})^i}{(\ib{u} - \ib{z})^2}
\end{equation}
is the Weizs\"acker-Williams soft gluon emission kernel. 
The $L$ and $R$ are ``left'' and ``right'' Lie derivatives\footnote{The naming of the derivatives may seem counter-intuitive, but they appear on the opposite side to what is expected due to the lightcone time axis running from right to left in our diagrammatic notation.} that act to color-rotate the Wilson lines on the left and right sides of the target field, respectively. 
They are defined as
\begin{align}
\ld{a}{u} &\equiv - i g (\uu{u}{} t^a)_{\alpha\beta} \frac{\delta}{\delta \uu{u}{,\alpha\beta}},
\label{eq:defL}
\\
\rd{a}{u} &\equiv - i g (t^a \uu{u}{})_{\alpha\beta} \frac{\delta}{\delta \uu{u}{,\alpha\beta}}
\label{eq:defR}
\end{align}
where $\alpha,\beta$ are matrix indices and $\frac{\delta}{\delta \uu{u}{}}$ acts as an ordinary functional derivative:
\begin{align}
\frac{\delta}{\delta \uu{u}{,\alpha\beta}} \uu{x}{,\gamma\rho} 
= \delta_{\alpha\gamma} \delta_{\beta\rho} \delta^{(2)}(\ib{u} - \ib{x})
\equiv \delta_{\alpha\gamma} \delta_{\beta\rho} \delta_{\ib{ux}}.
\end{align}

We can represent the action of the Lie derivatives on the Wilson lines as
\begin{align}
\rd{a}{u} \uu{x}{} &= - i g \delta_\ib{ux} t^a \uu{x}{}
= - i g \overset{a}{\diag{glu}} \diag{targref} \diag{empty}, 
\label{act1}
\\
\ld{a}{u} \uu{x}{} &= - i g \delta_\ib{ux} \uu{x}{} t^a
= - i g \diag{empty} \diag{targref} \overset{a}{\diag{glu}}.
\rule{0pt}{10ex} 
\end{align}
The Hermitian conjugates of these expressions give
\begin{align}
\rd{a}{u} \ud{x}{} &= i g \delta_\ib{ux} \ud{x}{} t^a
= i g \overset{a}{\diag{glu}} \diag{targ} \diag{empty},
\\
\ld{a}{u} \ud{x}{} &= i g \delta_\ib{ux} t^a \ud{x}{}
= i g \diag{empty} \diag{targ} \overset{a}{\diag{glu}}.
\rule{0pt}{10ex} 
\label{act2}
\end{align}
The left and right Lie derivatives $L$ and $R$ are related to each other by
\begin{align}
\ld{a}{u} &= \ua{\dagger ab}{u}{} \rd{b}{u} 
\\
\diag{empty} \diag{targref} \overset{a}{\diag{glu}}
\; &= \overset{b}{\diag{gluout}} \diag{targgluref} \diag{emptyglu_a},
\end{align}
which follows from the identity $\ua{\dagger ab}{u}{} t^b = \uu{u}{} t^a \ud{u}{}$.
The Lie derivatives satisfy the commutation relations 
\begin{align}
[ \ld{a}{u}, \ld{b}{v} ] &= g \delta_\ib{uv} f^{abc} \ld{c}{u}, 
\\
[ \rd{a}{u}, \rd{b}{v} ] &= - g \delta_\ib{uv} f^{abc} \rd{c}{u}, 
\\
[ \ld{a}{u}, \rd{b}{v} ] &= [ \ld{a}{u}, \ua{\dagger ab}{z}{} ] = [ \rd{a}{u}, \ua{\dagger ab}{z}{} ] = 0. 
\end{align}

\subsection{The JIMWLK equation in the Langevin formalism}

Generically, it can be shown that a Fokker-Planck description of a system's dynamics can be recast in a Langevin description. For the JIMWLK equation, this was done in \cite{Blaizot:2002xy}, with a slight simplification introduced in \cite{Lappi:2012vw} using the form of \eq\nr{ham} with left and right derivatives. Numerical solutions to JIMWLK evolution are more conveniently expressed using the Langevin formulation of the equation, as opposed to the Fokker-Planck formulation discussed in the previous section. This is one reason to explore analytically the Langevin picture. 
In this formulation, evolution is treated as a random walk in the functional space of Wilson lines. 
Rapidity acts as a ``time'' and is discretized as $Y - Y_0 = \epsilon N$ with $\mathbb{Z} \ni N \rightarrow \infty, \epsilon \rightarrow 0$, where each evolution step is labelled by $n \in \{0,1,...,N\}$. The averaging over the probability distribution of Wilson lines in \eq\nr{exp-val} is equivalent to the averaging over a noise term in the stochastic equation. 
This term can be taken as a localized Gaussian white noise
\begin{align}
\left\langle \nu^{i,a}_{\ib{x},m} \nu^{j,b}_{\ib{y},n} \right\rangle = \frac{1}{\epsilon} \delta^{ij} \delta^{ab} \delta_{mn} \delta_\ib{xy},
\label{noise}
\end{align}
where $\nu^{i,a}_{\ib{z},m} \in \mathbb{R}$.
The noise is introduced within terms we can call, respectively, ``left'' and ``right'' (traceless, Hermitian) color fields
\begin{align}
\al{x}{n}
& \equiv \frac{1}{\sqrt{4\pi^3}} \int_\ib{z} \mathcal{K}^i_\ib{xz} \nu^i_{\ib{z},n}, 
\\
\ar{x}{n} 
& \equiv \frac{1}{\sqrt{4\pi^3}} \int_\ib{z} \mathcal{K}^i_\ib{xz} \uu{z}{,n} \nu^i_{\ib{z},n} \ud{z}{,n},
\label{eq:defar}
\end{align}
where $\nu^i_{\ib{z},n} \equiv \nu^{i,a}_{\ib{z},n} t^a$ is an element of the SU$(\nc)$ algebra.

These definitions of $\alpha^L$ and $\alpha^R$ can be interchanged, as long as one is rotated by $\ua{\dagger ab}{}{}$ with respect to the other; this merely amounts to a redefinition of the noise in \eq\nr{noise} \cite{Lappi:2012vw}. To be explicit, one defines a rotated noise as
\begin{equation}
\tilde{\nu}^i_{\ib{z},n} \equiv \tilde{\nu}^{i,a}_{\ib{z},n} t^a \equiv \uu{z}{,n} \nu^i_{\ib{z},n} \ud{z}{,n} .
\label{eq:rotnoise}
\end{equation}
It is then straightforward to show that this rotated noise is also a Gaussian random variable with 
\begin{align}
\left\langle \tilde{\nu}^{i,a}_{\ib{x},m} \tilde{\nu}^{jb}_{\ib{y},n} \right\rangle = \frac{1}{\epsilon} \delta^{ij} \delta^{ab} \delta_{mn} \delta_\ib{xy}.
\label{nunu}
\end{align}
With this, one can equally well consider the rotated noise $\tilde{\nu}$ as being independent of the Wilson lines. 
Thus, the original noise $\nu$, and so too $\alpha^L$, are quantities that depend on them. However, the choice of whether to consider $\nu$ or $\tilde{\nu}$ has to be made globally for the whole calculation at once, and kept fixed when taking functional derivatives with respect to the Wilson lines. These properties will be crucial later to see how the evolution between the two rapidities becomes independent of the Wilson lines. 

The Langevin equation describing the evolution of a Wilson line from step $n$ to step $n+1$ is written as
\begin{align}
\ud{x}{,n+1} = \epl{x}{n} \ud{x}{,n} \emr{x}{n}.
\label{wilson}
\end{align}
The two matrix exponentials act as infinitesimal color rotations to the left and to the right of the target field, hence the naming of the $\alpha$'s. 
If the Wilson line at rapidity step $n$ is represented as
\begin{align}
\ud{x}{,n} \equiv \diag{empty} \diag{targ} \diag{empty},
\end{align}
then we can write
\begin{align}
\al{x}{n} \ud{x}{,n} &= \frac{1}{\sqrt{4\pi^3}} \int_\ib{z} \mathcal{K}^i_\ib{xz} \nu^{i,a}_{\ib{z},n} \diag{empty} \diag{targ} \overset{a}{\diag{glu}} \diag{x},
\\
\ar{x}{n} \ud{x}{,n} &= \frac{1}{\sqrt{4\pi^3}} \int_\ib{z} \mathcal{K}^i_\ib{xz} \nu^{i,a}_{\ib{z},n} \overset{b}{\diag{gluout}} \diag{targglu} \diag{emptyglu_a} \diag{xz}.
\end{align}
Notice that the free color index in these two diagrams contracts with the color index of the noise factor.
The Hermiticity of the color fields means that the time step for the complex conjugate Wilson line is simply
\begin{align}
\uu{x}{,n+1} = \epr{x}{n} \uu{x}{,n} \eml{x}{n}.
\label{wilson-hc}
\end{align}

\subsubsection{Expansion in the time step $\epsilon$} 

Since $\epsilon$ is infinitesimal, we may choose to keep terms only up to order $\epsilon$. 
The color fields each contain a factor $\nu$, which is of order $\epsilon^{-1/2}$, and always appear multiplied by a factor of $\epsilon$.
So $\mathcal{O}(\epsilon \alpha^{L/R}) = \mathcal{O}(\epsilon^{1/2})$ and we can immediately neglect from the expansion powers of $\alpha^{L/R}$ larger than two. Also, since in the end we only need to keep terms of order $(\alpha^{L/R})^2$ multiplied by terms that do not depend on the noise, we can at any stage in the calculation take the expectation value over the noise in such terms.

It is useful to note that Wilson lines appear only in the cross term $\alpha^{L}\alpha^{R}$, but not in the squares:
\begin{equation}
\langle (\al{x}{n})^2 \rangle
= 
\langle (\ar{x}{n})^2 \rangle
\\
= \frac{C_F}{4\pi^3\epsilon} \int_\ib{z} \mathcal{K}_\ib{xxz}.
\label{alphal}
\end{equation}
In terms of diagrams, this can be expressed as the relation
\begin{equation}
\label{eq:lrvirtual}
\diag{empty} \diag{empty} \diag{targ} \overset{a}{\diag{gluout}} \overset{a}{\diag{gluin}}
= 
\overset{a}{\diag{gluout}} \overset{a}{\diag{gluin}} \diag{targ} \diag{empty} \diag{empty}.
\end{equation}
Using these, one step of the Wilson line to the order needed can be written as 
\begin{align}
 \ud{x}{,n+1} \!\!\!\!\!\!\! 
 & \nonumber \\
=& \; \ud{x}{,n} 
+ \int_\ib{z} \left( \frac{i \epsilon g}{\sqrt{4\pi^3}} \mathcal{K}^i_\ib{xz} \nu^{i,a}_{\ib{z},n} - \frac{\epsilon g^2}{4\pi^3} \mathcal{K}_\ib{xxz} t^a \right) \nonumber
\\
& \times
(t^a \ud{x}{,n} - \ud{x}{,n} \ua{\dagger ab}{z}{,n} t^b)
+ \mathcal{O}(\epsilon^{3/2}) 
\label{ud-eps}
\\
=& \; \diag{empty} \diag{targ} \diag{empty} \diag{x} \nonumber
\\
& + \frac{i \epsilon g}{\sqrt{4\pi^3}} \int_\ib{z} \mathcal{K}^i_\ib{xz} \nu^{i,a}_{\ib{z},n} \left( \diag{empty} \diag{targ} \overset{a}{\diag{glu}} \diag{x}
\!\!\!\!\!\!
- \overset{b}{\diag{gluout}} \diag{targglu} \diag{emptyglu_a} \diag{xz} 
\!\!\!\!\!\!\!\!\!
\right) \nonumber
\\ 
& - \frac{\epsilon g^2}{4\pi^3} \int_\ib{z} \mathcal{K}_\ib{xxz} \left( 
\diag{targ} \overset{a}{\diag{gluout}} \overset{a}{\diag{gluin}} \diag{x}
\!\!\!\!\!\!
- \overset{b}{\diag{gluout}} \diag{targglu} \overset{a}{\diag{gluin}} \diag{xz}
\!\!\!
\right) \nonumber
\\
& + \mathcal{O}(\epsilon^{3/2}),
\end{align}
where the relation \nr{eq:lrvirtual} can be used to draw the virtual diagram on either side of the target.

By calculating the dipole $\hat{S}_{\ib{x\bar{x}},n+1}$ in terms of quantities at the previous step $n$, we can also derive the first equation of the Balitsky hierarchy:
\begin{multline}\label{eq:balitsky1}
\frac{d}{dY} \frac{\tr{\ud{x}{} \uu{y}{}}}{\nc}
= - \frac{\nc}{2} \frac{\alpha_s}{\pi^2} \int_\ib{z} \tilde{\mathcal{K}}_\ib{xyz} \left(
\frac{\tr{\ud{x}{,n} \uu{y}{,n}}}{\nc} \right.
\\
- \left. \frac{\tr{\ud{z}{,n} \uu{y}{,n}}}{\nc} \frac{\tr{\ud{x}{,n} \uu{z}{,n}}}{\nc} \right)
+ \mathcal{O}(\epsilon^{3/2}).
\end{multline}
Here, one needs to use the Fierz identity
\begin{multline}
2 \ua{ab}{z}{} \tr{\ud{y}{} t^a \uu{x}{} t^b} = 
\tr{\ud{z}{} \uu{x}{}} \tr{\ud{y}{} \uu{z}{}} 
\\
- \frac{1}{\nc} \tr{\ud{y}{} \uu{x}{}}
\end{multline}
to simplify the color structure and get rid of the adjoint representation matrices. 
Note that at this order in $\epsilon$ (and thus also in the limit $\epsilon \to 0$), \eq\nr{eq:balitsky1} is exact, even at the level of a single configuration without expectation values. 
Additionally, taking the expectation value on both sides and using the mean field approximation 
$
\left< \tr{\ud{z}{,n} \uu{y}{,n}} \tr{\ud{x}{,n}\uu{z}{,n}} \right>
\approx 
\left< \tr{\ud{z}{,n} \uu{y}{,n}}
\right> \left<
\tr{\ud{x}{,n} \uu{z}{,n}} \right>,
$
transforms \eq\nr{eq:balitsky1} into the Balitsky-Kovchegov (BK) equation
\begin{multline}
\frac{d}{dY} \left< \frac{\tr{\ud{x}{} \uu{y}{}}}{\nc} \right> =
\\
- \frac{\nc}{2} \frac{\alpha_s}{\pi^2} \int_\ib{z} \tilde{\mathcal{K}}_\ib{xyz} \left(
\left< \frac{\tr{\ud{x}{,n} \uu{y}{,n}}}{\nc} \right> \right.
\\
- \left. \left< \frac{\tr{\ud{z}{,n} \uu{y}{,n}}}{\nc} \right> \left< \frac{\tr{\ud{x}{,n} \uu{z}{,n}}}{\nc} \right> \right)
+ \mathcal{O}(\epsilon^{3/2}).
\label{eq:bk}
\end{multline}

\section{One- and two-particle production at equal rapidity}
\label{sec:equal}

Consider a single quark produced in a proton-nucleus collision. It is described mathematically by a fundamental representation dipole
\begin{align}
\hat{S}_\ib{x\bar{x}} \equiv \frac{\tr{U^\dagger_\ib{x} \bar{U}_\ib{\bar{x}}}}{\nc}
= \frac{1}{\nc} \tr{\diag{empty_x} \diag{targ} \diag{cut} \diag{targ} \diag{empty_xb}},
\label{dipole}
\end{align}
where the dashed line denotes the separation between the direct amplitude (DA) on the left and the complex conjugate amplitude (CCA) on the right. 
The bars on both the Wilson line and the coordinate in $\bar{U}_\ib{\bar{x}}$ denote that this Wilson line is in the CCA. 
Henceforth, this bar notation will be used to distinguish between quantities in the DA (unbarred) and the CCA (barred). 
The cross section for inclusive quark production in a proton-nucleus collision is then 
\begin{align}
\frac{d\sigma_{q}}{d\eta_p d^2\ib{p}} = x q (x) \frac{1}{(2 \pi)^2} \int_\ib{x\bar{x}} e^{-i \ib{p} \cdot (\ib{x} - \ib{\bar{x}})} \left\langle \left. \hat{S}_\ib{x\bar{x}} \right|_{\bar{U} = U} \right\rangle_Y.
\end{align}
Here, $Y$ is the relative rapidity of the produced quark with respect to the target, $x$ is the longitudinal momentum fraction of the projectile, $x q(x)$ is the quark distribution in the proton, and $\ib{p}$ and $\eta_p$ are the transverse momentum and rapidity, respectively, of the quark. 
The dipole expectation value can be obtained by averaging the dipole operator over the probability distribution of the Wilson lines according to \eq\nr{exp-val}, except that at this stage, the DA and CCA should still be regarded as independent:
\begin{align}
\left\langle \hat{S}_\ib{x\bar{x}} \right\rangle_Y = \int [DU] [D\bar{U}] W_Y[U,\bar{U}] \hat{S}_\ib{x\bar{x}}.
\label{eq-rap}
\end{align}

For inclusive quark-gluon production (both the quark and the gluon emitted with rapidity $Y$ with respect to the target), the cross section can be written compactly in terms of a ``production Hamiltonian'' \cite{Kovner:2006ge,Kovner:2006wr,Iancu:2013uva} operating on the quark cross section:
\begin{multline}
\frac{d\sigma_{qg}}{d\eta_p d^2\ib{p} \, d\eta_k d^2\ib{k}} = 
\\
\frac{1}{(2 \pi)^4} \int_\ib{x\bar{x}} e^{-i \ib{p} \cdot (\ib{x} - \ib{\bar{x}})} \left\langle \left. H_\mathrm{prod}(\ib{k}) \hat{S}_\ib{x\bar{x}} \right|_{\bar{U} = U} \right\rangle_Y.
\label{cross-sec}
\end{multline}
Here, the quark has transverse momentum $\ib{p}$ and pseudo-rapidity $\eta_p$, and the gluon has transverse momentum $\ib{k}$ and pseudo-rapidity $\eta_k$.
The production Hamiltonian is given by~\cite{Iancu:2013uva}
\begin{multline}
H_\mathrm{prod} (\ib{k}) = \frac{1}{4\pi^3} \int _\ib{y \bar{y}} e^{-i \ib{k} \cdot (\ib{y} - \ib{\bar{y}})} \int_\ib{u\bar{u}} \mathcal{K}^i_\ib{yu} \mathcal{K}^i_\ib{\bar{y} \bar{u}}
\\ \times
(\ld{a}{u} - \ua{\dagger ab}{y}{} \rd{b}{u}) (\ldb{a}{u} - \uab{\dagger ac}{y}{} \rdb{c}{u}),
\label{ham-bar}
\end{multline}
which correctly accounts for all possible ways that a second gluon can be produced. 
Notice that transverse coordinates $\ib{y}$ and $\ib{\bar{y}}$ are kept distinct, and the Lie derivatives with respect to the Wilson lines in the DA and CCA are kept separate. In spite of the notational similarity, this makes $H_\mathrm{prod}$ a somewhat more complicated operator than the JIMWLK Hamiltonian in \eq\nr{ham}.

In order to evaluate the cross section, the four terms in
\begin{align}
(\ld{a}{u} - \ua{\dagger ab}{y}{} \rd{b}{u}) (\ldb{a}{u} - \uab{\dagger ac}{y}{} \rdb{c}{u}) \hat{S}_\ib{x\bar{x}}
\label{four-terms}
\end{align}
need to be calculated, where the left and right Lie derivatives can be evaluated using \eqs\nr{act1} and~\nr{act2}.
Only once all the functional derivatives have been evaluated can we set $\bar{U} = U,$ since there is no more need to distinguish between Wilson lines in the DA and CCA separately. Substituting the results into \eq\nr{cross-sec} gives
\begin{multline}
\frac{d\sigma_{2g}}{d\eta_p d^2\ib{p} \, d\eta_k d^2\ib{k}} = 
\\
\frac{1}{(2 \pi)^4} \frac{\as}{\pi^2} \int_\ib{x\bar{x} y \bar{y}} e^{-i \ib{p} \cdot (\ib{x} - \ib{\bar{x}}) - i \ib{k} \cdot (\ib{y} - \ib{\bar{y}})} \mathcal{K}^i_\ib{yx} \mathcal{K}^i_\ib{\bar{y} \bar{x}} 
\\ \times \left\langle C_f \frac{\tr{\uu{\bar{x}}{} \ud{x}{}}}{\nc}
- (\ua{\dagger ab}{\bar{y}}{} + \ua{\dagger ab}{y}{}) \frac{\tr{t^b \uu{\bar{x}}{} t^a \ud{x}{}}}{\nc} \right.
\\
+ \left. (\uu{\bar{y}}{} \ud{y}{})^{ab} \frac{\tr{t^a \uu{\bar{x}}{} \ud{x}{} t^b}}{\nc} \right\rangle_Y,
\label{eq:laidet}
\end{multline}
where the integrals over $\ib{u}$ and $\ib{\bar{u}}$ have been evaluated using the delta functions $\delta_\ib{\bar{u}\bar{x}} \delta_\ib{ux}$ from calculating the Lie derivatives, and the overall constant has been rewritten using $\as = g^2/(4\pi)$. 
This is the analogue of the result (in the soft gluon limit $z\to 0$) of two-gluon production at equal rapidity obtained in Ref.~\cite{Iancu:2013dta}. Note that the rapidities of the quark and gluon are not really equal, because in the production Hamiltonian one has taken the limit where the gluon is soft. However, at this point they are also not parametrically large, so that one would need to consider high energy evolution between the two. Thus, the rapidity separation in \eq\nr{eq:laidet} $\Delta Y$ satisfies $1 \ll \Delta Y \ll 1/\as$.

\section{Dilute limit: stochastic picture of BFKL evolution}
\label{sec:dilute}

In order to understand the connection of the Langevin picture of JIMWLK evolution to the physics of the BFKL equation, we must develop the Wilson lines in the limit of a small color field. To do this, we start with the fundamental representation Wilson line
\begin{align}
\ud{x}{,n} 
& \equiv P \exp \left\{ i g \int dx^+ \alpha^a_{\ib{x},n}(x^+) t^a \right\} \nonumber
\\
& \equiv e^{i \la{x}{,n}{}} \nonumber
\\
& = \mathbb{1} + i \la{x}{,n}{} - \frac{1}{2} \la{x}{,n}{2} + \mathcal{O}(\la{}{}{3}),
\end{align}
where each real matrix $\lambda$ is an element of the algebra of SU$(\nc)$ and denotes a one-gluon interaction between projectile and target. 
We can represent this diagrammatically as
\begin{align}
\ud{x}{,n} = \diag{empty} + i \diag{lam} - \frac{1}{2} \diag{lam} \diag{lam} + \mathcal{O}(\la{}{}{3}).
\end{align}
The Hermitian conjugate is simply
\begin{align}
\uu{x}{,n} 
= e^{-i \la{x}{,n}{}} 
= \mathbb{1} - i \la{x}{,n}{} - \frac{1}{2}\la{x}{,n}{2} + \mathcal{O}(\la{}{}{3})
\end{align}
and an adjoint Wilson line is
\begin{multline}
\ua{\dagger ab}{x}{,n} 
= \delta^{ab} + f^{abc} \la{x}{,n}{c} + 2 \tr{t^a \la{x}{,n}{} t^b \la{x}{,n}{}} 
\\
- \tr{t^a \{ t^b, \la{x}{,n}{2} \}} + \mathcal{O}(\la{}{}{3}).
\end{multline}

The part of the Langevin step for Wilson lines in \eq\nr{ud-eps} involving an adjoint representation Wilson line corresponds to the interaction of a gluon with the target shockwave. At step $n=0$ in this linearized limit, we have 
\begin{multline}
\ud{x}{,0} \ua{\dagger ab}{z}{,0} t^b
= t^a - f^{abc} \la{z}{,0}{b} t^c + i\la{x}{,0}{} t^a + \mathcal{O}(\la{}{}{2}),
\end{multline}
which we can represent as
\begin{multline}
\overset{b}{\diag{gluout}} \diag{targglu} \diag{emptyglu_a} \diag{xz} = 
\diag{empty} \overset{a}{\diag{glu}} \diag{empty} 
- \overset{c}{\diag{gluout}} \underset{\ib{z}}{\diag{lam2_b}} \diag{emptyglu_a}
\\
+ i \overset{a}{\diag{glu}} \underset{\ib{x}}{\diag{lam}} \diag{empty}
+ \mathcal{O}(\la{}{}{2}).
\end{multline}
Using this, we can write the full Langevin step to linear order in the gluon field as
\begin{multline}
\la{x}{,n+1}{}
= \la{x}{,n}{}
+ \int_\ib{z} \left( \frac{i \epsilon g}{\sqrt{4\pi^3}} \mathcal{K}^i_\ib{xz} \nu^{i,a}_{\ib{z},n} - \frac{\epsilon g^2}{4\pi^3} \mathcal{K}_\ib{xxz} t^a \right) 
\\
\times i f^{abc} t^c (\la{x}{,n}{b} - \la{z}{,n}{b})
+ \mathcal{O}(\epsilon^{3/2}, \la{}{}{2}).
\label{langevin-dilute}
\end{multline}
Diagrammatically, we have
\begin{multline}
\la{x}{,n+1}{} = \diag{empty}\underset{\ib{x}}{\diag{lam}}\diag{empty}
\\
+ \frac{i \epsilon g}{\sqrt{4\pi^3}} \int_\ib{z} \mathcal{K}^i_\ib{xz} \nu^{i,a}_{\ib{z},n} i
\left( \overset{c}{\diag{gluout}} \underset{\ib{x}}{\diag{lam2_b}} \diag{emptyglu_a}
- \overset{c}{\diag{gluout}} \underset{\ib{z}}{\diag{lam2_b}} \diag{emptyglu_a} \right)
\\
- \frac{\epsilon g^2}{4\pi^3} \int_\ib{z} \mathcal{K}_\ib{xxz} 
\left( \overset{c}{\diag{gluout}} \underset{\ib{x}}{\diag{lam2_b}} \overset{\;a}{\diag{gluin}}
- \overset{c}{\diag{gluout}} \underset{\ib{z}}{\diag{lam2_b}} \overset{a}{\diag{gluin}} \right)
\\
+ \mathcal{O}(\epsilon^{3/2}, \la{}{}{2}).
\end{multline}
Note that the 3-gluon vertex can also be written as a commutator:
\begin{align}
i f^{abc} t^c \la{x}{,n}{b} 
& = [t^a, \la{x}{,n}{}] \\
i \overset{c}{\diag{gluout}} \underset{\ib{x}}{\diag{lam2_b}} \diag{emptyglu_a}
\; & = \overset{a}{\diag{glu}} \underset{\ib{x}}{\diag{lam}} \diag{empty}
- \diag{empty} \underset{\ib{x}}{\diag{lam}} \overset{a}{\diag{glu}}.
\end{align}

\eq\nr{langevin-dilute} is simply a linear iterative equation, for which we can write a formal solution. 
Separating out a time evolution matrix $\mathcal{M}$ as 
\begin{multline}
\mathcal{M}^{ab}_{\ib{xw},n}
\equiv \delta_\ib{xw} \delta^{ab}
\\
+ \int_\ib{z} \left( \frac{\epsilon g}{\sqrt{4\pi^3}} \mathcal{K}^i_\ib{xz} \nu^{i,c}_{\ib{z},n} f^{abc}
- \frac{\epsilon g^2}{4\pi^3} \frac{\nc}{2} \mathcal{K}_\ib{xxz} \delta^{ab} \right) 
\\ \times
\left(\delta_\ib{xw} - \delta_\ib{zw} \right),
\label{eq:defM}
\end{multline}
we can write 
\begin{align}
\la{x}{,n+1}{a}
= \int_\ib{w} \mathcal{M}^{ab}_{\ib{xw},n} \la{w}{,n}{b}
+ \mathcal{O}(\epsilon^{3/2}, \la{}{}{2}).
\label{laelement}
\end{align}
This recursive relation has the solution
\begin{multline}
\la{x}{,n+1}{a}
= \int_\ib{w_n} \mathcal{M}^{ab_n}_{\ib{xw_n},n} 
\prod_{j=0}^{n-1} \left( \int_\ib{w_j} \mathcal{M}^{b_{j+1}b_j}_{\ib{w_{j+1}w_j},j} \right)
\la{w_\mathrm{0}}{,0}{b_0}.
\label{recur}
\end{multline}
The product of $\mathcal{M}$'s can be simplified further if necessary, by using \eq\nr{noise} and keeping terms up to linear order in $\epsilon$.

\subsubsection{Reggeization}

As a side note, we can now easily see (following \cite{Caron-Huot:2013fea}) how the gluon field $\lambda$ ``Reggeizes'' in this picture. This is done by taking the expectation value of the single gluon field time step \eq\nr{langevin-dilute}, which eliminates terms linear in $\nu$ and leads to 
\begin{multline}
\frac{1}{\epsilon} \langle \la{x}{,n+1}{} - \la{x}{,n}{} \rangle = 
\\
- \frac{\nc}{2} \frac{g^2}{4\pi^3} \int_\ib{z} \mathcal{K}_\ib{xxz}
\langle \la{x}{,n}{} - \la{z}{,n}{} \rangle 
+ \mathcal{O}(\epsilon^{2}, \la{}{}{2}),
\end{multline}
which we may write as
\begin{multline}
\left\langle \frac{d}{dY} \la{x}{}{} \right\rangle
= \frac{\nc}{2} \frac{\alpha_s}{\pi^2} \int_\ib{z} \mathcal{K}_\ib{xxz}
\langle \la{z}{,n}{} - \la{x}{,n}{} \rangle + \mathcal{O}(\epsilon^{2}, \la{}{}{2}).
\label{lam-evol}
\end{multline}
This equation can be Fourier transformed using
\begin{align}
\la{}{}{a}(\ib{p}) = \int_\ib{z} e^{i \ib{p \cdot z}} \la{z}{}{a}
\end{align}
and written as 
\begin{align}
\left\langle \frac{d}{dY} \la{}{}{a}(\ib{p}) \right\rangle 
= \langle \alpha_g(\ib{p}) \la{}{n}{a}(\ib{p}) \rangle + \mathcal{O}(\epsilon^{2}, \la{}{}{2}).
\end{align}
Here, 
\begin{align}
\alpha_g(\ib{p}) \equiv \frac{\nc}{2} \frac{\alpha_s}{\pi^2} \int_\ib{z} \frac{1}{\ib{z}^2} (e^{i \ib{p \cdot z}} - 1)
\end{align}
is the so-called ``Regge trajectory'' (see \cite{Caron-Huot:2013fea}). The gluon is said to ``Reggeize'', meaning that the amplitude for the gluon exchange process scales with energy to the power of the trajectory. 
In the Langevin picture, Reggeization therefore simply refers to the power law growth of the expectation value of the gluon field.

\subsubsection{The BFKL equation}

Let us then obtain the BFKL equation for the unintegrated gluon distribution. This can be done by first expanding the Wilson lines in the dilute limit, and looking at the evolution of a quantity that is quadratic in the expansion parameter $\lambda$. For this discussion, it is useful to still keep track of the $\lambda$'s in the DA and CCA separately. 
In order to derive such a quadratic equation, we square \eq\nr{laelement}.
Eventually taking the expectation value of the noise on both sides will remove terms linear in $\nu$ and simplify terms quadratic in $\nu$ according to \eq\nr{nunu}. As discussed above, we have expanded to the relevant order in the rapidity step $\epsilon$. After some straightforward color algebra, we obtain
\begin{multline}
\la{x}{,n+1}{a} \lab{x}{,n+1}{a}
= \; \int_\ib{w\bar{w}} \bigg[ \delta_\ib{wx} \delta_\ib{\bar{w}\bar{x}}
\\
- \frac{\nc}{2} \frac{\epsilon \alpha_s}{\pi^2} \int_\ib{z} (\mathcal{K}_\ib{\bar{x}\bar{x}z} \delta_\ib{wx}
(\delta_\ib{\bar{w}\bar{x}} - \delta_\ib{\bar{w}z})
+ \mathcal{K}_\ib{xxz}
(\delta_\ib{wx} - \delta_\ib{wz}) \delta_\ib{\bar{w}\bar{x}}
\\
- 2 \mathcal{K}_\ib{x\bar{x}z}
(\delta_\ib{wx} - \delta_\ib{wz})
(\delta_\ib{\bar{w}\bar{x}} - \delta_\ib{\bar{w}z})) \bigg] \la{w}{,n}{a} \lab{w}{,n}{a} + \mathcal{O}(\epsilon^{3/2}, \la{}{}{3}).
\label{eq:lamlamnonoise}
\end{multline}

From this basic equation, one can define two different versions of the BFKL equation. The first one comes naturally when one considers a particle production process, in which one wants to keep a nonzero contribution from both the DA and the CCA. To do this, we define the unintegrated gluon distribution
\begin{equation}
\phi^n_{\ib{x\bar{x}}} \equiv \langle \la{x}{,n}{a} \lab{x}{,n}{a} \rangle.
\label{eq:deffi}
\end{equation}
Then, \eq\nr{eq:lamlamnonoise} yields the evolution equation 
\begin{multline}
\phi^{n+1}_{\ib{x\bar{x}}} - \phi^n_{\ib{x\bar{x}}}
= - \frac{\nc}{2} \frac{\epsilon \alpha_s}{\pi^2} \int_\ib{z} [\mathcal{K}_\ib{xxz} (\phi^n_{\ib{x\bar{x}}} - \phi^n_{\ib{z\bar{x}}})
\\
+ \mathcal{K}_\ib{\bar{x}\bar{x}z} (\phi^n_{\ib{x\bar{x}}} - \phi^n_{\ib{xz}})
- 2 \mathcal{K}_\ib{x\bar{x}z} (\phi^n_{\ib{x\bar{x}}} - \phi^n_{\ib{xz}} - \phi^n_{\ib{z\bar{x}}} + \phi^n_{\ib{zz}})]
\\
+ \mathcal{O}(\epsilon^{3/2}, \phi^{3/2}).
\label{eq:textbookbfkl}
\end{multline}
Note that this does not have the customary form of a coordinate space BFKL equation. Equation~\nr{eq:textbookbfkl}  acquires a more familiar form, however, in momentum space. To see this, we define the Fourier transform
\begin{equation}
 \phi_\ib{x\bar{x}} \equiv \int\frac{d^2\ib{p}}{(2\pi)^2}
 e^{-i\ib{p} \cdot (\ib{x-\bar{x}})} \phi(\ib{p}) 
 \equiv
\int_{\ib{p}}
 e^{-i\ib{p} \cdot (\ib{x-\bar{x}}) }\phi(\ib{p}).
 \end{equation}
$\phi_\ib{x\bar{x}}$ is a function of the relative length $\ib{x - \bar{x}}$, so we need not transform each coordinate separately.
Using this and the Fourier transformed kernel
\begin{equation}
 \mathcal{K}^i_\ib{uv} = 
2\pi i
 \int_{\ib{k}}
e^{- i\ib{k} \cdot (\ib{u-v}) }\frac{k^i}{\ib{k}^2},
\label{kernelft}
\end{equation}
the BFKL equation \nr{eq:textbookbfkl} becomes
\begin{multline}
\phi^{n+1}({\ib{q}})
= \phi^n({\ib{q}})
\\
+ 4 \nc \epsilon \alpha_s \int_\ib{p} \frac{1}{\ib{(q - p)}^2} \bigg(
\frac{\phi^n({\ib{p}}) \ib{p}^2}{\ib{q}^2}
- \frac{1}{2} \frac{\phi^n({\ib{q}}) \ib{q}^2}{\ib{p}^2} \bigg)
\\
+ \mathcal{O}(\epsilon^{3/2}, \phi^{3/2}).
\label{eq:bfklft}
\end{multline}
This is immediately recognized as the (color singlet, zero momentum transfer) textbook version of the BFKL equation \cite{Forshaw:1997dc}. 

The other (Mueller's) version of the BFKL equation~\cite{Mueller:1993rr,Mueller:1994jq} is obtained when we set the Wilson lines to be equal in the DA and CCA. One then looks at the expansion of the dipole operator up to order $\lambda^2$ as
\begin{align}
\frac{\tr{\ud{x}{} \uu{y}{}}}{\nc}
&= 1 - \frac{1}{4 \nc} (\la{x}{}{a} - \la{y}{}{a}) (\la{x}{}{a} - \la{y}{}{a}) + \mathcal{O}(\la{}{}{3}).
\label{lamdipole}
\end{align}
The natural definition of the gluon distribution based on the expansion of the dipole operator is then the so-called ``BFKL pomeron'' \cite{Caron-Huot:2013fea}
\begin{equation}
\varphi_\ib{xy} \equiv \left< (\la{x}{}{a} - \la{y}{}{a}) (\la{x}{}{a} - \la{y}{}{a})\right>,
\label{eq:defugd}
\end{equation}
which we can write in terms of $\phi$ by setting $\bar{\lambda} = \la{}{}{}$:
\begin{equation}
\phi_\ib{xx}+ \phi_\ib{yy} - 2\phi_\ib{xy} \overset{\bar{\lambda} = \la{}{}{}}{=} \varphi_\ib{xy}.
\label{varphitophi}
\end{equation}
Using \eq\nr{eq:textbookbfkl} for each of the three terms in \eq\nr{varphitophi}, one arrives at the Mueller version of the BFKL equation:
\begin{multline}
\varphi^{n+1}_{\ib{xy}} - \varphi^n_{\ib{xy}} =
\\
- \frac{\nc}{2} \frac{\epsilon \alpha_s}{\pi^2} \int_\ib{z} 
 \tilde{\mathcal{K}}_\ib{xyz}
[\varphi^n_{\ib{xy}} - \varphi^n_{\ib{xz}} - \varphi^n_{\ib{zy}}].
\label{eq:bfklphi}
\end{multline}
This is the version of the BFKL equation one usually sees written in coordinate space. It can also be obtained by linearizing the BK equation \nr{eq:bk}.

It is important to emphasize the relatively trivial but important observation that the two equations, \nr{eq:textbookbfkl} and \nr{eq:bfklphi}, are closely related but not the same. \eq\nr{eq:textbookbfkl} is usually derived and written in momentum space by considering the BFKL ladder diagrams. It appears naturally for a particle production process in which we want to have an explicit product of the DA and the CCA. Such a calculation would begin with the dipole written diagrammatically as
\begin{align*}
\frac{\tr{\ud{x}{} \uub{x}{}}}{\nc}
= \frac{1}{\nc} \tr{\diag{empty_x} \diag{targ} \diag{cut} \diag{targ} \diag{empty_xb}}.
\end{align*}
On the other hand, the Mueller version of the BFKL equation \nr{eq:bfklphi}, is satisfied by an object associated with the elastic amplitude for a color neutral dipole. In our notation, this would be expressed as
\begin{align*}
\frac{\tr{\ud{x}{} \uu{y}{}}}{\nc}
= \frac{1}{\nc} \tr{\diag{empty2} \diag{targ2} \diag{empty2_xy}},
\end{align*}
which is not naturally separated into terms associated with the DA and the CCA. While \eq\nr{eq:bfklphi} is usually written in coordinate space, the momentum space version is straightforward to obtain; it is not particularly simple nor useful in this context, so we omit it here.

\section{Unequal rapidity correlators in JIMWLK}
\label{sec:neqy}

\subsection{Two-particle production}

Next, we consider the production of two particles produced at parametrically different rapidities. The rapidity $Y_A$ is closer to that of the target, i.e. it is ``earlier'' on the evolution trajectory of the target. We want to calculate the double inclusive cross section for the simultaneous production of another particle at a later rapidity $Y$ that is much larger than the first one: $\as(Y-Y_A) \gg 1$. Unlike the case considered in Sec.~\ref{sec:equal}, we now have genuine high energy evolution between the two produced particles due to the rapidity separation. The situation is rendered more complicated than the JIMWLK evolution considered in 
Sec.~\ref{sec:jimwlk} by the fact that the Wilson line trajectories are now \emph{conditional} ones. This is because they are now aware of the fact that a particle with a specific transverse momentum was produced earlier in the evolution. The IT Langevin formalism for this scenario was set up in \cite{Iancu:2013uva}. The purpose of our discussion here is to elucidate how this formalism in the dilute limit relates to a conventional BFKL picture. 

We begin this discussion as formulated in \cite{Iancu:2013uva}. After the first part of evolution from rapidity $Y_\mathrm{in}$ to $Y_A$ for the quark dipole, one obtains the Wilson lines $\bar{U}_{\ib{\bar{x}},A}$ and $U^\dagger_{\ib{x},A}$ at the earlier rapidity $Y_A$. In order to keep track of the gluon that is produced at $Y_A$, we consider these Wilson lines to be fixed for now. 
They act as the initial condition for the second part of the evolution from rapidity $Y_A$ to $Y$. In terms of the discretization $Y-Y_A = N\epsilon$, we have $Y_0 \equiv Y_A$, $U_0 \equiv \uu{}{A}$ and $\bar{U}_0 \equiv \bar{U}_A$ at $n=0$.
The expectation value of the cross section for producing a quark at some rapidity $Y$ is then calculated as an average over the noise $\nu$ at the end of the stochastic process:
\begin{align}
\left\langle \hat{S}_{\ib{x\bar{x}}} \right\rangle_{Y-Y_A} = \left\langle \hat{S}_{\ib{x\bar{x}},N} \right\rangle_\nu.
\label{langevin-ave}
\end{align}
\eq\nr{exp-val} 
for the expectation value of an operator at the later rapidity $Y$ is now written as
\begin{align}
\left\langle \hat{\mathcal{O}} 
\right\rangle_{Y - Y_A} 
\equiv \int [DU D\bar{U}] W_{Y - Y_A} [U, \bar{U}| \uu{}{A}, \bar{U}_A] \hat{\mathcal{O}}. 
\label{a-dipole}
\end{align}

We need to have a new conditional weight function $W_{Y - Y_A} [U, \bar{U}| \uu{}{A}, \bar{U}_A]$ \cite{Gelis:2008sz}, which is the probability of observing Wilson lines $U$ and $\bar{U}$ at rapidity $Y$, with the condition that there are already Wilson lines $\uu{}{A}$ and $\bar{U}_A$ at the earlier rapidity $Y_A$. This weight function obeys the differential equation 
\begin{align}
\frac{\partial}{\partial Y} W_{Y - Y_A} [U, \bar{U}| \uu{}{A}, \bar{U}_A] = H_\mathrm{evol} W_{Y - Y_A} [U, \bar{U}| \uu{}{A}, \bar{U}_A]
\end{align}
(cf. \eq\nr{differential-eq}).
The evolution Hamiltonian $H_\mathrm{evol}$ is just the conventional JIMWLK Hamiltonian, with the exception that one must now keep track of the Wilson lines and Lie derivatives for the DA and the CCA separately. Thus, there are terms operating only on the DA (11), terms operating only on the CCA (22) and a mixed term (12):
\begin{align}
H_\mathrm{evol} \equiv H_{11} + 2H_{12} + H_{22},
\end{align}
where 
\begin{align}
H_{11} \equiv \frac{1}{8 \pi^3} \int _\ib{uvz} \mathcal{K}_\ib{uvz} (\ldalt{a}{u}{n} - \ua{\dagger ab}{z}{,n} \rdalt{b}{u}{n}) \nonumber
\\
\times (\ldalt{a}{v}{n} - \ua{\dagger ac}{z}{,n} \rdalt{c}{v}{n}), 
\label{eq:hevol11}
\\
H_{12} \equiv \frac{1}{8 \pi^3} \int _\ib{u\bar{v}z} \mathcal{K}_\ib{u\bar{v}z} (\ldalt{a}{u}{n} - \ua{\dagger ab}{z}{,n} \rdalt{b}{u}{n}) \nonumber
\\
\times (\ldbalt{a}{v}{n} - \bar{\tilde{U}}^{\dagger ac}_{\ib{z},n} \rdbalt{c}{v}{n}), 
\label{eq:hevol12}
\\
H_{22} \equiv \frac{1}{8 \pi^3} \int _\ib{\bar{u}\bar{v}\bar{z}} \mathcal{K}_\ib{\bar{u}\bar{v}\bar{z}} (\ldbalt{a}{u}{n} - \uab{\dagger ac}{z}{,n} \rdbalt{b}{u}{n}) \nonumber
\\
\times (\ldbalt{a}{v}{n} - \uab{\dagger ac}{z}{,n} \rdbalt{c}{v}{n})
\label{eq:hevol22}
\end{align}
(cf. \eq\nr{ham} or rather, \nr{ham-bar}).
Here, the Lie derivatives and the adjoint Wilson lines carry a subscript $n$ to emphasize that they pertain to the current (latest) rapidity in the evolution.

The initial condition at $Y_A$ for the conditional weight function sets Wilson lines for both the DA and the CCA:
\begin{align}
W_{Y_A} [U, \bar{U}| \uu{}{A}, \bar{U}_A] = \delta [U - \uu{}{A}] \delta [\bar{U} - \bar{U}_A].
\end{align}
If one sets the DA and the CCA to be the same at the initial condition, this property is preserved throughout the evolution; at some other rapidity $Y_B$, we then have
\begin{align}
W_{Y_B} [U, \bar{U}| \uu{}{A}, \uu{}{A}] = \delta [U - \bar{U}] W_{Y_B} [U| \uu{}{A}],
\end{align}
where $W_{Y_B} [U| \uu{}{A}]$ is just the conventional conditional probability (e.g. \eq\nr{differential-eq}) with initial condition 
\begin{align}
W_{Y_A} [U| \uu{}{A}] = \delta [U - \uu{}{A}].
\end{align}

The cross section for inclusive two-particle production, as provided by \eq\nr{cross-sec}, needs to be modified to account for the fact that the emitted particle has a different rapidity to the projectile.
For a gluon emitted from a quark projectile at rapidity $Y_A$, we must now operate with the production Hamiltonian \nr{ham-bar} \emph{acting on the Wilson lines at} $Y_A$, i.e. on the initial condition for the conditional evolution \cite{Iancu:2013uva}:
\begin{multline}
\frac{d\sigma_{qg}}{dY d^2\ib{p} \, dY_A d^2\ib{k}_A}
= \frac{1}{(2 \pi)^4} \int_\ib{x\bar{x}} e^{-i \ib{p} \cdot (\ib{x} - \ib{\bar{x}})}
\\
\times \left\langle \left. H_\mathrm{prod}(\ib{k}_A)
\left\langle \hat{S}_{\ib{x\bar{x}}} \right\rangle_{Y - Y_A} \right|_{\bar{U}_A = \uu{}{A}} \right\rangle_{Y_A}.
\label{uneq-cs}
\end{multline}
Note that there are now two separate averages. The dipole operator is evolved from $Y_A$ to $Y$ with the conditional JIMWLK evolution; the average over these Langevin trajectories is denoted by $\langle\rangle_{Y - Y_A}$. One then operates with the production Hamiltonian, which is a functional derivative with respect to the initial conditions $U_A$ and $\bar{U}_A$. Only then does one set the Wilson lines to be the same in the DA and the CCA. Finally, one does the average over the earlier rapidity distribution $\langle\rangle_{Y_A}$, using a weight function that has been evolved from $Y_\mathrm{in}$ to $Y_A$.
This double averaging procedure can be written in a more explicit way with the help of delta functions and conditional probabilities:
\begin{multline}
\frac{d\sigma_{qg}}{dY d^2\ib{p} \, dY_A d^2\ib{k}_A}
= \frac{1}{(2 \pi)^4} 
\int_\ib{x\bar{x}} e^{-i \ib{p} \cdot (\ib{x} - \ib{\bar{x}})} 
\\
\times \int [D\uu{}{A}] W_{Y_A}[\uu{}{A}] \int [D\bar{U}_A] \delta [\bar{U}_A - \uu{}{A}] H_\mathrm{prod}(\ib{k}_A) 
\\
\times \int [DU D\bar{U}] 
 W_{Y - Y_A} [U, \bar{U}| \uu{}{A}, \bar{U}_A] \hat{S}_{\ib{x\bar{x}}}.
\label{eq:xs}
 \end{multline}

In order to proceed, we need the result of the production Hamiltonian operating on the dipole, i.e. we need to calculate $H_\mathrm{prod} \hat{S}_{\ib{x}\ib{\bar{x}}}$ . This expression will have several terms, with left or right Lie derivatives operating on the Wilson lines. We have to also maintain the distinction between barred and unbarred contributions. The resulting expressions not only involve Wilson lines at the rapidity $Y_A \equiv Y_0$, but also at later rapidities $Y_n$, where the $n$ refers to the discrete Langevin process description we are using. Putting everything together, we can write the cross section as
\begin{multline}
\frac{d\sigma_{qg}}{dY d^2\ib{p} \, dY_A d^2\ib{k}_A}
= \frac{1}{(2 \pi)^4} \frac{1}{4\pi^3} \frac{1}{\nc} \int_\ib{x\bar{x}y\bar{y}} e^{-i \ib{p} \cdot (\ib{x} - \ib{\bar{x}})} 
\\
\times e^{-i\ib{k}_A \cdot (\ib{y} - \ib{\bar{y}})} 
\int_\ib{uv} 
\mathcal{K}^i_\ib{yu} \mathcal{K}^i_\ib{\bar{y}v} \left\langle \left\langle \mathcal{I}_N \right\rangle_\ib{\nu} \right\rangle_{Y_A},
\label{final} 
\end{multline}
where
\begin{multline}
\mathcal{I}_n := \tr{\ldbalt{a}{u}{0} \uub{x}{,n} \ldalt{a}{u}{0} \ud{x}{,n}}
\\
- \uab{\dagger ac}{y}{,0} \tr{\rdbalt{c}{u}{0} \uub{x}{,n} \ldalt{a}{u}{0} \ud{x}{,n}}
\\
- \ua{\dagger ab}{y}{,0} \tr{\ldbalt{a}{u}{0} \uub{x}{,n} \rdalt{b}{u}{0} \ud{x}{,n}}
\\
+ \ua{\dagger ab}{y}{,0} \uab{\dagger ac}{y}{,0} \tr{\rdbalt{c}{u}{0} \uub{x}{,n} \rdalt{b}{u}{0} \ud{x}{,n}}.
\label{mathcal-i}
\end{multline}
Before moving forward, let us stress some features of \eqs\nr{final} and~\nr{mathcal-i}.  In addition to the Wilson lines at rapidity $Y_N$, they involve Lie derivatives of Wilson lines at $Y_N$ with respect to Wilson lines at $Y_A$, e.g. $\ldbalt{a}{u}{0} \uub{x}{,N}$, denoted by the subscript $0$ in the Lie derivatives. This is contrary to the evolution Hamiltonian in \eqs\nr{eq:hevol11}, \nr{eq:hevol12}, \nr{eq:hevol22} that involved Lie derivatives w.r.t. the latest rapidity in the evolution. Additionally, the expression contains adjoint representation Wilson lines at the initial rapidity $Y_A$, again denoted by the subscript $0$. These Lie derivatives are new degrees of freedom, in some sense like Reggeized gluon propagators, that encode information about the (de)correlation in rapidity of the Wilson lines. The JIMWLK equation also gives the evolution equations for the Lie derivatives themselves. To find the expressions for $R\ud{}{}, R\uu{}{}, L\ud{}{}$ and $L\uu{}{}$, one starts by acting with the Lie derivatives on \eqs\nr{wilson} and \nr{wilson-hc}. However, the four equations are not independent of each other. For example, we may start by finding the equation for $R\ud{}{}$. The Hermitian conjugate will give the equation for $R\uu{}{}$, and the relation $\ldalt{a}{u}{0} = \ua{\dagger ab}{u}{,0} \rdalt{b}{u}{0}$ can be used to get the equations for $L\ud{}{}$ and $L\uu{}{}$.

The initial conditions at $Y_0$ for these four bilocal Langevin equations are given by \eqs\nr{act1} to \nr{act2}:
\begin{eqnarray}
\label{eq:ic1}
\rdalt{a}{u}{n} \ud{x}{,0} &=& i g \delta_\ib{ux} \ud{x}{,0} t^a,
\\
\label{eq:ic2}
\rdalt{a}{u}{0} \uu{x}{,0} &=& - i g \delta_\ib{ux} t^a \uu{x}{,0},
\\
\label{eq:ic3}
\ldalt{a}{u}{0} \ud{x}{,0} &=& i g \delta_\ib{ux} t^a \ud{x}{,0},
\\
\label{eq:ic4}
\ldalt{a}{u}{0} \uu{x}{,0} &=& - i g \delta_\ib{ux} \uu{x}{,0} t^a.
\end{eqnarray}
We emphasize once more that the adjoint Wilson lines appearing in $\mathcal{I}_n$ are from the production Hamiltonian; they are always at $Y_0$ and do not evolve to $Y_N$.

\subsection{Evolution in rapidity}

To derive the evolution equations for the Lie derivatives, we begin by acting on \eq\nr{wilson} with a right Lie derivative. After some manipulations described in \cite{Iancu:2013uva}, we arrive at
\begin{multline}
\rdalt{a}{u}{0} \ud{x}{,n+1}
= \epl{x}{n} \rdalt{a}{u}{0} \ud{x}{,n} \emr{x}{n}
\\
- \frac{i \epsilon g}{\sqrt{4\pi^3}} \epl{x}{n} \ud{x}{,n} \int_\ib{z} \mathcal{K}^i_\ib{xz}
\\
\times [\uu{z}{,n} \nu^i_{\ib{z},n} \ud{z}{,n}, \uu{z}{,n} \rdalt{a}{u}{0} \ud{z}{,n}].
\label{ru-dagger}
\end{multline}
The quantity $\rdalt{a}{u}{0} \ud{x}{,n}$, describes the evolution from the initial rapidity $Y_A$, initial coordinate $\ib{u}$ and color index $a$, to the final Wilson line at rapidity step $n$, coordinate $\ib{z}$ and a color that is encoded in the matrix structure of $\rdalt{a}{u}{0} \ud{z}{,n}$. 

In order to understand what \eq\nr{ru-dagger} does at each iteration, we can look at one step in evolution in diagrams. 
First, we Taylor expand the exponentials in $\epsilon$. 
At $n=0$, we can use the initial condition \eq\nr{eq:ic1} to write
\begin{multline}
\rdalt{a}{u}{0} \ud{x}{,1} 
= ig \delta_\ib{ux} \ud{x}{,0} t^a
\\
+ ig \int_\ib{z} \left( \frac{i \epsilon g}{\sqrt{4\pi^3}} \mathcal{K}^i_\ib{xz} \nu^{i,b}_{\ib{z},0} - \frac{\epsilon g^2}{4\pi^3} \mathcal{K}_\ib{xxz} t^b \right) 
\\
\times [\delta_\ib{ux} (t^b \ud{x}{,0} t^a 
- \ud{x}{,0} t^a \ua{\dagger bc}{z}{,0} t^c)
\\
- \delta_\ib{uz} (\ud{x}{,0} \ua{\dagger bc}{z}{,0} t^c t^a
- \ud{x}{,0} t^a \ua{\dagger bc}{z}{,0} t^c)]
+ \mathcal{O}(\epsilon^{3/2}).
\end{multline}
In diagrams, this is
\begin{widetext}
\begin{multline}
\rdalt{a}{u}{0} \ud{x}{,1} 
= ig \overset{a}{\diag{glu}} \diag{targ} \diag{empty} \diag{xu}
\\
+ ig \frac{i \epsilon g}{\sqrt{4\pi^3}} \int_\ib{z} \mathcal{K}^i_\ib{xz} \nu^{i,b}_{\ib{z},0}
\left( \overset{a}{\diag{glu}} \diag{targ} \overset{b}{\diag{glu}} \diag{xu}
- \overset{c}{\diag{gluoutlong}} \overset{a}{\diag{glu2glu}} \diag{targ2glu} \diag{empty2glu_b} \diag{xuz}
- \overset{a}{\diag{glu}} \overset{c}{\diag{gluout}} \diag{targglu} \diag{emptyglu_b} \diag{xzu}
+ \overset{c}{\diag{gluoutlong}} \overset{a}{\diag{glu2glu}} \diag{targ2glu} \diag{empty2glu_b} \diag{xzu2} \right)
\\
- ig \frac{\epsilon g^2}{4\pi^3} \int_\ib{z} \mathcal{K}_\ib{xxz}
\left( \overset{a}{\diag{glu}} \diag{targ} \overset{b}{\diag{gluout}} \overset{b}{\diag{gluin}} \diag{xu}
- \overset{c}{\diag{gluoutlong}} \overset{a}{\diag{glu2glu}} \diag{targ2glu} \overset{b}{\diag{gluinlong}} \diag{xuz}
- \overset{a}{\diag{glu}} \overset{c}{\diag{gluout}} \diag{targglu} \overset{b}{\diag{gluin}} \diag{xzu}
+ \overset{c}{\diag{gluoutlong}} \overset{a}{\diag{glu2glu}} \diag{targ2glu} \overset{b}{\diag{gluinlong}} \diag{xzu2} \right)
+ \mathcal{O}(\epsilon^{3/2}).
\label{ru-dagger-diags}
\end{multline}
\end{widetext}
From this, we can see that after one iteration, \eq\nr{ru-dagger} contains all possible diagrams with one gluon inserted on the right of the target (corresponding to color index $a$ in the diagrams), and afterwards either none or one more gluon insertion. The second gluon (corresponding to color indices $b$ and $c$) can be inserted in four different ways: either to the right or to the left of the target, and either at coordinate $\ib{x}$ or $\ib{z}$. Additionally, it can be either left as a free color index (contracted by the color of the noise), or reabsorbed by the quark. This makes the $2^3$ two-gluon diagrams shown in \eq\nr{ru-dagger-diags}. 

The next iteration of the evolution equation will add more gluons to the diagrams present in \eq\nr{ru-dagger-diags}, in the same way. The number of terms therefore grows very rapidly with each step. The same analysis can be done for the $L\ud{}{}$ evolution. The only difference will be that the gluons appear on the opposite side of the target.

It is  more natural to write \eq\nr{ru-dagger} such that the color structure is more explicit. The propagator should have two adjoint representation indices at the ends, i.e. $R^{ab}$, or equivalently, be an explicitly Hermitian traceless matrix with one additional index, i.e. $R^{ab}t^b$. This more symmetric form is achieved by multiplying $R\ud{}{}$ by $\uu{}{}$ from the left. Defining $R^a_{\ib{ux},n} \equiv \uu{x}{,n} \rdalt{a}{u}{0} \ud{x}{,n}$, which is conveniently a member of the Lie algebra of SU($\nc$), we can write the Langevin step compactly as
\begin{multline}
R^a_{\ib{ux},n+1} = \epr{x}{n} R^a_{\ib{ux},n} \emr{x}{n}
\\ 
- \frac{i \epsilon g}{\sqrt{4\pi^3}} \epr{x}{n} \int_\ib{z} \mathcal{K}^i_\ib{xz} [\tilde{\nu}^i_{\ib{z},n}, R^a_{\ib{uz},n}],
\label{ru-dagger2}
\end{multline}
where we have used $\tilde{\nu}^i_{\ib{z},n} = \uu{z}{,n} \nu^i_{\ib{z},n} \ud{z}{,n}$ as introduced in \eq\nr{eq:rotnoise}. Notice that this equation is linear in $R^a_{\ib{ux},n}$, although it resums all orders in the background field, incorporated into the Wilson line.

As mentioned above, the three other required equations, $R\uu{}{}, L\ud{}{}$ and $L\uu{}{}$, can be obtained directly from \eq\nr{ru-dagger2}. In terms of explicitly Hermitian quantities,  we obtain for $\bar{R}^a_{\ib{ux},n} \equiv (\rdalt{a}{u}{0} \uu{x}{,n} )\ud{x}{,n}$:
\begin{multline}
\bar{R}^a_{\ib{ux},n+1}
= \; \epr{x}{n}\bar{R}^a_{\ib{ux},n} \emr{x}{n}
\\
- \frac{i \epsilon g}{\sqrt{4\pi^3}} \int_\ib{z} \mathcal{K}^i_\ib{xz} 
[\tilde{\nu}^i_{\ib{z},n}, 
 \bar{R}^a_{\ib{uz},n}] \emr{x}{n}.
\label{ru2} 
\end{multline}
Defining
$L^a_{\ib{ux},n} \equiv \uu{x}{,n} \ldalt{a}{u}{0} \ud{x}{,n}$,
we get the corresponding equations for the left Lie derivative:
\begin{multline}
L^a_{\ib{ux},n+1}
= \epr{x}{n} L^a_{\ib{ux},n} \emr{x}{n}
\\ - \frac{i \epsilon g}{\sqrt{4\pi^3}} \epr{x}{n} \int_\ib{z} \mathcal{K}^i_\ib{xz} [\tilde{\nu}^i_{\ib{z},n}, 
L^a_{\ib{uz},n}]
\label{lu-dagger2_orig}
\end{multline}
and for $\bar{L}^a_{\ib{ux},n} \equiv (\ldalt{a}{u}{0} \uu{x}{,n})\ud{x}{,n}$,
\begin{multline}
\bar{L}^a_{\ib{ux},n+1}
= \epr{x}{n} \bar{L}^a_{\ib{ux},n} 
\emr{x}{n}
\\
\quad- \frac{i \epsilon g}{\sqrt{4\pi^3}} \int_\ib{z} \mathcal{K}^i_\ib{xz} [\tilde{\nu}^i_{\ib{z},n}, 
\bar{L}^a_{\ib{uz},n}] 
\emr{x}{n}.
\label{lu2_orig}
\end{multline}
The initial conditions for this set of four evolution equations follow directly from \eqs\nr{eq:ic1} to \nr{eq:ic4}:
\begin{align}
R^a_{\ib{ux},0} &= i g \delta_\ib{ux} t^a,
\label{ic-new1}
\\
\bar{R}^a_{\ib{ux},0} &= - i g \delta_\ib{ux} t^a,
\\
L^a_{\ib{ux},0} &= i g \delta_\ib{ux} \ua{\dagger ab}{x}{,0} t^b,
\label{ic-new3}
\\
\bar{L}^a_{\ib{ux},0} &= - i g \delta_\ib{ux} \ua{\dagger ab}{x}{,0} t^b.
\label{ic-new4}
\end{align}

A quick inspection of the equations of motion \nr{ru-dagger2} to \nr{lu2_orig} reveals in all of them two crucial features. 
Firstly, they only depend on the ``rotated noise'' $\tilde{\nu}$ (see \eq\nr{eq:rotnoise}), but not the original unrotated noise $\nu$; note that this is true also for the ``right'' color field $\ar{x}{n}$, \eq\nr{eq:defar}. Secondly, their dependence on the Wilson line also comes through the rotated noise, but not separately in terms of explicit Wilson lines in the evolution equations. Thus, if we take the rotated noise as the independent variable that one averages over, all the dependence on the Wilson lines disappears. This means that the quantities $R^a_{\ib{ux},n}$, $\bar{R}^a_{\ib{ux},n}$, $L^a_{\ib{ux},n}$ and $\bar{L}^a_{\ib{ux},n}$ satisfy evolution equations that are linear and independent of the Wilson lines, and we can therefore express the evolution between the two rapidities in terms of linear BFKL-like dynamics. 
This can be made even more explicit by developing the equations in $\epsilon$ and, as usual, replacing the terms that are quadratic in the noise with their expectation values. Doing this one gets
\begin{multline}
R^a_{\ib{ux},n+1} =
R^a_{\ib{ux},n}
+ \frac{i \epsilon g}{\sqrt{4\pi^3}} \int_\ib{z} \mathcal{K}^i_\ib{xz} [\tilde{\nu}^{i}_{\ib{z},n}, R^a_{\ib{ux},n} - R^a_{\ib{uz},n}] 
\\
- \frac{N_c}{2} \frac{\epsilon g^2}{4\pi^3} \int_\ib{z} \mathcal{K}_\ib{xxz} (R^a_{\ib{ux},n} - R^a_{\ib{uz},n}) 
+ \mathcal{O}(\epsilon^{3/2}),
\label{ru-dagger2-new}
\end{multline}
\begin{multline}
\bar{R}^a_{\ib{ux},n+1} =
\bar{R}^a_{\ib{ux},n}
+ \frac{i \epsilon g}{\sqrt{4\pi^3}} \int_\ib{z} \mathcal{K}^i_\ib{xz} [\tilde{\nu}^{i}_{\ib{z},n}, \bar{R}^a_{\ib{ux},n} - \bar{R}^a_{\ib{uz},n}]
\\
- \frac{N_c}{2} \frac{\epsilon g^2}{4\pi^3} \int_\ib{z} \mathcal{K}_\ib{xxz} (\bar{R}^a_{\ib{ux},n} - \bar{R}^a_{\ib{uz},n}) 
+ \mathcal{O}(\epsilon^{3/2}),
\label{ru2-new} 
\end{multline}
\begin{multline}
L^a_{\ib{ux},n+1} =
L^a_{\ib{ux},n}
+ \frac{i \epsilon g}{\sqrt{4\pi^3}} \int_\ib{z} \mathcal{K}^i_\ib{xz} [\tilde{\nu}^{i}_{\ib{z},n}, L^a_{\ib{ux},n} - L^a_{\ib{uz},n}] 
\\
- \frac{N_c}{2} \frac{\epsilon g^2}{4\pi^3} \int_\ib{z} \mathcal{K}_\ib{xxz} (L^a_{\ib{ux},n} - L^a_{\ib{uz},n}) 
+ \mathcal{O}(\epsilon^{3/2}),
\label{lu-dagger2_orig-new}
\end{multline}
\begin{multline}
\bar{L}^a_{\ib{ux},n+1} =
\bar{L}^a_{\ib{ux},n}
+ \frac{i \epsilon g}{\sqrt{4\pi^3}} \int_\ib{z} \mathcal{K}^i_\ib{xz} [\tilde{\nu}^{i}_{\ib{z},n}, \bar{L}^a_{\ib{ux},n} - \bar{L}^a_{\ib{uz},n}]
\\
- \frac{N_c}{2} \frac{\epsilon g^2}{4\pi^3} \int_\ib{z} \mathcal{K}_\ib{xxz} (\bar{L}^a_{\ib{ux},n} - \bar{L}^a_{\ib{uz},n}) 
+ \mathcal{O}(\epsilon^{3/2}).
\label{lu2_orig-new}
\end{multline}

The whole cross section, however, is not given by a ``$k_T$-factorized'' expression (unlike the dilute case that we discuss in the next section). This is not true even at equal rapidity, as we have seen in \eq\nr{eq:laidet}. This is due to the appearance of the Wilson lines in two places in the cross section. Firstly the initial conditions for $R^a_{\ib{ux},n}$, $\bar{R}^a_{\ib{ux},n}$, $L^a_{\ib{ux},n}$ and $\bar{L}^a_{\ib{ux},n}$ depend on the Wilson lines at $Y_A$. 
Secondly, the final expression for the cross section \nr{mathcal-i} written in terms of these quantities is now
\begin{multline}
\mathcal{I}_n 
= \tr{\bar{L}^{a}_{\ib{v\bar{x}},n} \uu{\bar{x}}{,n} \ud{x}{,n} L^a_{\ib{ux},n}}
\\
- \bar{\tilde{U}}^{\dagger ac}_{\ib{\bar{y}},0} \tr{\bar{R}^{c}_{\ib{v\bar{x}},n} \uu{\bar{x}}{,n} \ud{x}{,n} L^a_{\ib{ux},n}}
\\
- \ua{\dagger ab}{y}{,0} \tr{\bar{L}^{a}_{\ib{v\bar{x}},n} \uu{\bar{x}}{,n} \ud{x}{,n} R^b_{\ib{ux},n}}
\\
+ \ua{\dagger ab}{y}{,0} \bar{\tilde{U}}^{\dagger ac}_{\ib{\bar{y}},0} \tr{\bar{R}^{c}_{\ib{v\bar{x}},n} \uu{\bar{x}}{,n} \ud{x}{,n} R^b_{\ib{ux},n}}
\label{eq:icaln}
\end{multline}
We see that this expression involves explicit Wilson lines both at the rapidity of the earlier particle (i.e. $n=0$) and at the rapidity of the later one (i.e. $n$). Because of the latter, the expression in the full nonlinear case cannot be written in a $k_T$-factorized form with only an unintegrated gluon distribution at $n=0$ and a BFKL Green's function between the rapidities, as  we will do in the dilute limit in Sec.~\ref{sec:dilute2part}.

Because the evolution equations~\nr{ru-dagger2} to \nr{lu2_orig} do not depend on the Wilson lines, the dynamics of the evolution of the two-particle correlation between these rapidities is linear. This is a remarkable feature that should significantly simplify a future numerical analysis of the two-particle correlation in this framework. This is generically quite difficult in the original formulation due to the bilocal nature of the ``Reggeized gluon propagators'' $R^a_{\ib{ux},n}$, $\bar{R}^a_{\ib{ux},n}$, $L^a_{\ib{ux},n}$ and $\bar{L}^a_{\ib{ux},n}$ (i.e. the fact that they depend on two separate coordinate arguments). Note, however,  that calculating the  cross section requires additionally the nonlinear evolution of the Wilson lines themselves according to conventional JIMWLK. This evolution is \emph{correlated} with that of the Reggeized gluon propagators, since the evolution steps of both quantities are expressed in terms of the same noise at the same rapidity step. Since the evolution step of the latter is linear and does not explicitly involve the Wilson lines, it should be simpler to analyze. However, one \emph{cannot} factorize the expectation value of \eq\nr{eq:icaln} as a product of expectation values of a Wilson line operator at rapidity $n$ on one hand and a two-point function of the Reggeized gluon propagators on the other. 

Let us make, in passing, a side note concerning the initial conditions \nr{ic-new3} and \nr{ic-new4}. It would be tempting to get rid of the Wilson line in these initial conditions by defining, instead of
$L^a_{\ib{ux},n} \equiv \uu{x}{,n} \ldalt{a}{u}{0} \ud{x}{,n}$, a new quantity $ (\ldalt{a}{u}{0} \ud{x}{,n})\uu{x}{,n}$ and set out to solve its equation of motion. Contrary to $L^a_{\ib{ux},n}$, which has an initial condition that depends on the Wilson line but an evolution that can be expressed in a way that does not, this alternative quantity would have a simple initial condition (only a generator matrix $t^a$) but an evolution equation that depends on both the noise $\nu$ and the rotated noise $\tilde{\nu}$. This would make the linear dynamics of the evolution much more difficult to see explicitly.

The statement of linearity in the evolution between the two rapidities is in fact in agreement with the result for one quark and one gluon production in \cite{JalilianMarian:2004da}, where it is argued that there can be no pomeron mergings between these two particles. However, the situation becomes more complicated when one considers the production of one quark and two gluons. In this case, the result of \cite{JalilianMarian:2004da} is that evolution between the produced particles is genuinely nonlinear.

\section{Two-particle correlators in the dilute limit}
 \label{sec:dilute2part}

We shall now move on to work out what happens in the dilute limit, and show explicitly how to recover an expected BFKL result from this formalism. Our aim is to show that the two-particle cross section in the dilute limit can be expressed as a convolution of an unintegrated gluon distribution at the earlier rapidity $Y_A$, and a BFKL Green's function from $Y_A$ to the rapidity of the quark (see e.g. \cite{Leonidov:1999nc} for calculations of two-particle production from the same BFKL ladder). Although the following calculation is done by linearizing the evolution, it is important to note that due to the linearity of the dynamics in the full case discussed above, the nature of decorrelations in rapidity could be expected to be very similar. 

The general idea is to see what happens when we start increasing the rapidity separation between the produced quark and gluon. The gluon is produced at rapidity $Y_A$; the quark rapidity increases and the cross section changes accordingly, via JIMWLK evolution. It is important here to note that the JIMWLK evolution is that of the Wilson lines in the dipole operator $\hat{S}_{\ib{x\bar{x}}}$ at rapidity $Y_n$ only. The Wilson lines and Lie derivatives in the production Hamiltonian $H_\mathrm{prod}$ remain at the initial rapidity $Y_A$. In this sense, the JIMWLK evolution ``commutes'' with the production Hamiltonian and only operates on the dipole operator. We also know that the evolution must be linear, and thus in the dilute limit it must correspond to the BFKL equation \nr{eq:bfklft}. 

The essential part of the cross section \nr{final} is given by $\mathcal{I}_n$ as defined in \eq\nr{mathcal-i}: the combination of Wilson lines and their derivatives. To understand how this works, we have to understand the operations of the Lie derivatives in the dilute limit, i.e. how the Lie derivatives as defined in \eqs\nr{eq:defL} and \nr{eq:defR} act on $\lambda$. The result is obtained by changing the differentiation variable from elements of $U$ to $\lambda$, and then performing an expansion in powers of $\lambda$. Doing so gives
\begin{align}
\ldalt{a}{u}{0} &= g \left( \delta^{ac} - \frac{1}{2} f^{abc} \la{u}{,0}{b} + \mathcal{O}(\la{}{}{2}) \right) \frac{\delta}{\delta \la{u}{,0}{c}}, \nonumber
\\
\rdalt{a}{u}{0} &= g \left( \delta^{ac} + \frac{1}{2} f^{abc} \la{u}{,0}{b} + \mathcal{O}(\la{}{}{2}) \right) \frac{\delta}{\delta \la{u}{,0}{c}}
\label{eq:liederlambda}
\end{align}
(see also \cite{Caron-Huot:2013fea}).
As expected, the Lie derivatives reduce the term they act on by one power of $\lambda$; they are after all derivatives with respect to the Wilson line. Since the evolution equation for $\lambda$ is linear, this statement will be true at any step $n$. 

The most straightforward way to proceed would be to derive, in the spirit of~\cite{Iancu:2013uva}, evolution equations for the Lie derivatives operating on $\lambda$. This can be done in two equivalent ways: either by expanding the evolution equations of the bilocal quantities, e.g. \eq\nr{ru-dagger}, or by acting with the Lie derivative on the time step of $\lambda$ (\eq\nr{langevin-dilute}). In other words, the operations of Lie differentiating and expanding in $\lambda$ are commutative. Either procedure results in
\begin{multline}
\rdalt{a}{u}{0} \la{x}{,n+1}{}
= \rdalt{a}{u}{0} \la{x}{,n}{}
\\
+ \int_\ib{z} \left( \frac{i \epsilon g}{\sqrt{4\pi^3}} \mathcal{K}^i_\ib{xz} \nu^{i,d}_{\ib{z},n} - \frac{\epsilon g^2}{4\pi^3} \mathcal{K}_\ib{xxz} t^d \right) 
\\
\times i f^{dbc} t^c \rdalt{a}{u}{0} (\la{x}{,n}{b} - \la{z}{,n}{b})
+ \mathcal{O}(\epsilon^{3/2}, \la{}{}{2}).
\label{eq:rlambda}
\end{multline}
The equation for $L\ud{}{}$ is identical, with $R \rightarrow L$.
The initial conditions for the bilocal Langevin equations are given in \eqs\nr{eq:ic1} to \nr{eq:ic4}. By expanding them as
\begin{align}
\rdalt{a}{u}{0} \ud{x}{,0} &= i g \delta_\ib{ux} \left( 1 + i \la{x}{,0}{} - \frac{1}{2} \la{x}{,0}{2} \right) t^a + \mathcal{O}(\la{}{}{3}),
\\
\ldalt{a}{u}{0} \ud{x}{,0} &= i g \delta_\ib{ux} t^a \left(1 + i \la{x}{,0}{} - \frac{1}{2} \la{x}{,0}{2} \right) + \mathcal{O}(\la{}{}{3})
\end{align}
one easily obtains the initial conditions for the Lie derivatives of $\lambda$
as
\begin{equation}
\rdalt{a}{u}{0} \la{x}{,0}{} = \ldalt{a}{u}{0} \la{x}{,0}{} = gt^a \delta_\ib{ux}
+ \mathcal{O}(\la{}{}{}).
\end{equation}

In principle, we could go on to solve these equations, and indeed it is straightforward to write down a full formal solution in terms of the time evolution matrix $\mathcal{M}^{ab}_\ib{xy}$, as defined in \eq\nr{eq:defM}. But our main objective is to look at the cross section directly, and to derive an evolution equation for its dependence on the later rapidity $Y$. To do this, we use the expressions \nr{eq:liederlambda} to write the Lie derivatives in $H_\mathrm{prod}$ as 
\begin{multline}
(\ldbalt{a}{u}{0} - \uab{\dagger ac}{y}{,0} \rdbalt{c}{u}{0}) (\ldalt{a}{u}{0} - \ua{\dagger ab}{y}{,0} \rdalt{b}{u}{0}) =
\\
g^2 [f^{abc} f^{ade} (\lab{u}{,0}{e} - \lab{y}{,0}{e}) (\la{u}{,0}{c} - \la{y}{,0}{c})
+ \mathcal{O}(\la{}{}{3})] 
\\ \times
\frac{\delta}{\delta \lab{u}{,0}{d}} \frac{\delta}{\delta \la{u}{,0}{b}}.
\label{eq:linprodham}
\end{multline}
We emphasize here that one must take care to perform the functional derivatives in the right order. It is essential to operate first with the Lie derivatives in the DA and the CCA separately. Only after doing this is one allowed to set ${\bar{\lambda}}_0 = \lambda_0$ at the initial rapidity $Y_0 = Y_A$. On the other hand, the noise $\nu$ is the same in the DA and the CCA, so we do not need to keep them separated when averaging over the noise. For example, we can take the expectation value that allowed us to write \eq\nr{eq:lamlamnonoise}, even if the expression involves both $\lambda$ and $\bar{\lambda}$ separately.

An important feature of the operator \nr{eq:linprodham} is that it has exactly one derivative operating on the DA and one on the CCA. Also, since the DA and the CCA evolve separately, $\lambda_n$ depends only on $\lambda_0$ and not on $\bar{\lambda}_0$, and vice versa. Thus, from the expansion of the dipole in \eq\nr{eq:xs} as 
\begin{multline}
\hat{S}_{\ib{x}\ib{\bar{x}}} = 
1 - \frac{1}{4 \nc} \la{x}{}{a} \la{x}{}{a} - \frac{1}{4 \nc} \lab{x}{}{a} \lab{x}{}{a} + \frac{1}{2 \nc} \la{x}{}{a} \lab{x}{}{a} + \mathcal{O}(\la{}{}{3})
 \label{eq:lindipole}
\end{multline}
(cf. \eq\nr{lamdipole}), we need only to retain the cross term $\sim \lambda\bar{\lambda}$ when operating with the production Hamiltonian.

Using the linearized production Hamiltonian \nr{eq:linprodham} and the linearized dipole \nr{eq:lindipole}, we have
\begin{multline}
\mathcal{I}_n
= \frac{g^2}{2 \nc} f^{abc} f^{ade} (\lab{u}{,0}{e} - \lab{y}{,0}{e}) (\la{u}{,0}{c} - \la{y}{,0}{c}) 
\\
\times \frac{\delta}{\delta \lab{u}{,0}{d}} \frac{\delta}{\delta \la{u}{,0}{b}}
\lab{x}{,n}{f} \la{x}{,n}{f} + \mathcal{O}(\la{}{}{3}).
\end{multline}
Now we can write the explicit $\lambda$'s from the production Hamiltonian \nr{eq:linprodham} in terms of the gluon distribution $\phi^0_\ib{w\bar{w}}$ \nr{eq:deffi} at the initial rapidity $Y_A$. Then we can use $\delta^{ce}$ from the initial condition of $\phi$ to simplify the color algebra. 
We now introduce the definition
\begin{equation}
\mathcal{F}^{n}_{\ib{x,\bar{x},u,\bar{u}}} \equiv 
\frac{\delta}{\delta \lab{u}{,0}{a}} \frac{\delta}{\delta \la{u}{,0}{a}}
\lab{x}{,n}{b} \la{x}{,n}{b}
\label{eq:defFcal}
\end{equation}
for the BFKL Green's function. Note that due to the linear evolution of $\lambda$, the relation between $\la{x}{,n}{b}$ and $\la{u}{,0}{a}$ is linear (see e.g. \eq\nr{recur}). Thus the Green's function $\mathcal{F}^{n}_{\ib{x,\bar{x},u,\bar{u}}}$ defined by \eq\nr{eq:defFcal} does not depend on $\lambda$. Using this we get
\begin{multline}
\langle \mathcal{I}_N \rangle
= \frac{g^2}{2} (\phi^0_\ib{\bar{u} u} - \phi^0_\ib{\bar{u} y} - \phi^0_\ib{\bar{y} u} + \phi^0_\ib{\bar{y} y})
\mathcal{F}^N_\ib{x,\bar{x},u,\bar{u}} + \mathcal{O}(\phi^{3/2}),
\end{multline}
which we can put into the equation for the two-particle cross section \nr{eq:xs} to obtain a $k_T$-factorized expression:
\begin{multline}
\frac{d\sigma_{qg}}{dY d^2\ib{p} \, dY_A d^2\ib{k}_A} = 
\\
\frac{1}{(2 \pi)^4} \frac{1}{2 \nc} \frac{\alpha_s}{\pi^2} \int_\ib{x\bar{x}y\bar{y}u\bar{u}} 
\mathcal{K}^i_\ib{yu} \mathcal{K}^i_\ib{\bar{y}\bar{u}}
e^{-i \ib{p} \cdot (\ib{x} - \ib{\bar{x}}) -i\ib{k}_A \cdot (\ib{y} - \ib{\bar{y}})}
\\
\times (\phi^0_\ib{\bar{u} u} - \phi^0_\ib{\bar{u} y} - \phi^0_\ib{\bar{y} u} + \phi^0_\ib{\bar{y} y})
\mathcal{F}^N_\ib{x,\bar{x},u,\bar{u}} + \mathcal{O}(\phi^{3/2}).
\label{eq:cspfact}
\end{multline}

At this stage all the functional derivatives are already taken, and we can now finally take the color field to be equal in the DA and the CCA, i.e. take $\lambda = \bar{\lambda}$.  We can now, in fact, also replace the $\phi$'s with the other gluon distribution $\varphi$ \nr{eq:defugd}, since they appear in the particular combination 
\begin{multline}
\phi^0_\ib{\bar{u} u} - \phi^0_\ib{\bar{u} y} - \phi^0_\ib{\bar{y} u} + \phi^0_\ib{\bar{y} y}
\\= - \frac{1}{2} (\varphi^0_\ib{\bar{u} u} - \varphi^0_\ib{\bar{u} y} - \varphi^0_\ib{\bar{y} u} + \varphi^0_\ib{\bar{y} y}),
\end{multline}
where the value of the distribution at  zero coordinate separation cancels.
This means that the cross section can be equally well written in terms of the gluon distribution $\phi$ satisfying the usual BFKL equation \nr{eq:textbookbfkl} (more familiarly \eq\nr{eq:bfklft} in momentum space), or the BFKL pomeron $\varphi$ satisfying the Mueller version of the equation, \eq\nr{eq:bfklphi}. There is  a factor $-1/2$ difference due to our conventions here.  The fact that we are taking derivatives with respect to $\lab{u}{,0}{}$ and $\la{u}{,0}{}$ does not in any way interfere with the BFKL evolution of the gluon density $\lab{x}{,n}{b} \la{x}{,n}{b}$. So it is clear that $\mathcal{F}^{n}_{\ib{x,\bar{x},u,\bar{u}}}$ satisfies the same equation as $\lab{x}{,n}{b} \la{x}{,n}{b}$ with respect to the rapidity index $n$:
\begin{multline}
\mathcal{F}^{n+1}_{\ib{x,\bar{x},u,\bar{u}}} = \mathcal{F}^{n}_{\ib{x,\bar{x},u,\bar{u}}} 
\\
- \frac{\nc}{2} \frac{\epsilon \alpha_s}{\pi^2} \int_\ib{z} [\mathcal{K}_\ib{xxz} ( \mathcal{F}^{n}_{\ib{x,\bar{x},u,\bar{u}}} -  \mathcal{F}^{n}_{\ib{z,\bar{x},u,\bar{u}}})
\\
+ \mathcal{K}_\ib{\bar{x}\bar{x}z} ( \mathcal{F}^{n}_{\ib{x,\bar{x},u,\bar{u}}} - \mathcal{F}^{n}_{\ib{x,z,u,\bar{u}}})
\\
- 2 \mathcal{K}_\ib{x\bar{x}z} ( \mathcal{F}^{n}_{\ib{x,\bar{x},u,\bar{u}}} -  \mathcal{F}^{n}_{\ib{x,z,u,\bar{u}}} -  \mathcal{F}^{n}_{\ib{z,\bar{x},u,\bar{u}}} +  \mathcal{F}^{n}_{\ib{z,z,u,\bar{u}}})]
\\
+ \mathcal{O}(\mathcal{F}^{3/2})
\label{eq:bfklK}
\end{multline}
(cf. \eq\nr{eq:textbookbfkl}).

\begin{figure}
\centerline{\includegraphics[width=0.3\textwidth]{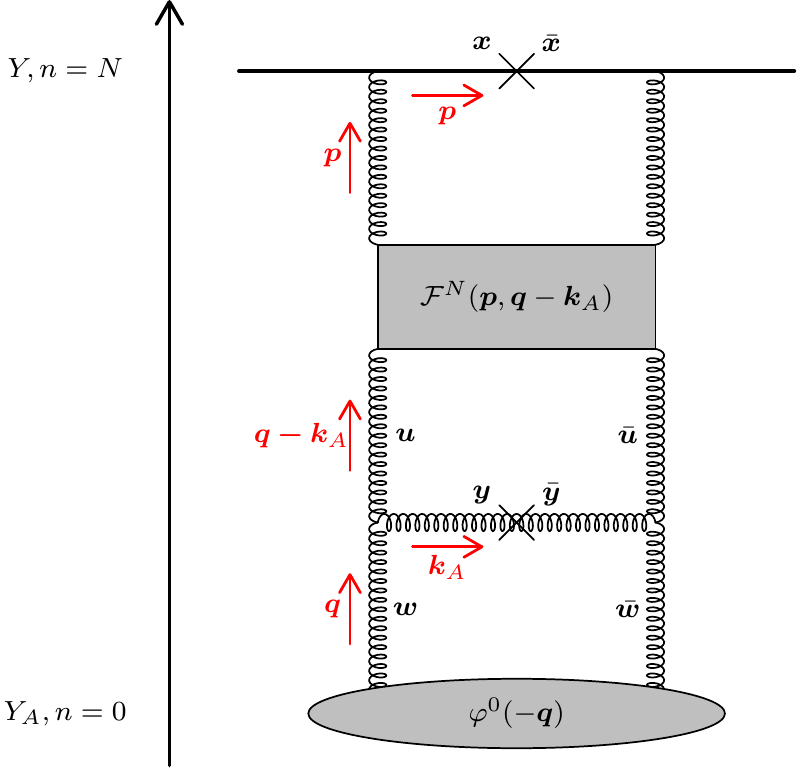}}
\caption{Coordinate and momentum assignments for the BFKL ladder diagram corresponding to \eq\nr{eq:fft}.}
\label{fig:ladder}
\end{figure}

Now that we have derived the factorized form for the cross section, \eq\nr{eq:cspfact}, it is easy to write it in momentum space.
We use the Fourier representations of the gluon emission kernels, given in \eq\nr{kernelft}, to write
\begin{multline}
\frac{d\sigma_{qg}}{dY d^2\ib{p} \, dY_A d^2\ib{k}_A} = 
\frac{1}{(2 \pi)^2} \frac{1}{4 \nc} \frac{\alpha_s}{\pi^2}  \int_\ib{x\bar{x}y\bar{y}u\bar{u}w\bar{w}l\bar{l}}
\frac{\ib{l\cdot \bar{l}}}{\ib{l}^2\ib{\bar{l}}^2}
\\
\times e^{- i \ib{l \cdot (y-u)} - i \ib{\bar{l} \cdot (\bar{y} - \bar{u}})}
e^{-i \ib{p} \cdot (\ib{x} - \ib{\bar{x}}) -i\ib{k}_A \cdot (\ib{y} - \ib{\bar{y}})}
\\
\times (\delta_\ib{wu} - \delta_\ib{wy}) (\delta_\ib{\bar{w}\bar{u}} - \delta_\ib{\bar{w}\bar{y}}) \varphi^0_\ib{w\bar{w}}
\mathcal{F}^N_\ib{x,\bar{x},u,\bar{u}} + \mathcal{O}(\varphi^{3/2}). 
\label{eq:ladderansatz}
\end{multline}
Introducing a Fourier representation for the BFKL Green's function
\begin{multline}
\mathcal{F}^{n}_{\ib{x,\bar{x},u,\bar{u}}} =
\\
\int_\ib{P\bar{P}m\bar{m}} 
e^{ - i (\ib{P \cdot x + \bar{P} \cdot \bar{x} + m \cdot u + \bar{m} \cdot \bar{u}})} 
\mathcal{F}^{n}(\ib{P,\bar{P},m,\bar{m}})
\end{multline}
and
\begin{equation}
\varphi_\ib{xy} \equiv \int_\ib{q} e^{ - i \ib{q} \cdot (\ib{x} - \ib{y})} \varphi(\ib{q}),
\end{equation}
we can write
\begin{multline}
\frac{d\sigma_{qg}}{dY d^2\ib{p} \, dY_A d^2\ib{k}_A} =
\\
- \frac{\alpha_s}{\nc} \int_\ib{q} \frac{\ib{q}^2}{(\ib{q} - \ib{k}_A)^2 \ib{k}_A^2}  \mathcal{F}^{N}(\ib{-p,p,\ib{q}-k_A,-q+k_A}) \varphi^0(-\ib{q})
\\ 
+ \mathcal{O}(\varphi^{3/2})
\label{eq:momspladder}
\end{multline}
with the (zero momentum transfer) BFKL Green's function $\mathcal{F}$ in momentum space satisfying the usual BFKL equation
\begin{multline}
\mathcal{F}^{n+1}(\ib{P},\ib{-P},\ib{m},\ib{-m}) = 
\mathcal{F}^{n}(\ib{P},\ib{-P},\ib{m},\ib{-m})
\\
+ 4 \nc \epsilon \alpha_s \int_\ib{K} \frac{1}{\ib{(P - K)}^2} \bigg( \mathcal{F}^{n}(\ib{K},\ib{-K},\ib{m},\ib{-m}) \frac{\ib{K}^2}{\ib{P}^2}
\\
- \frac{1}{2} \mathcal{F}^{n}(\ib{P},\ib{-P},\ib{m},\ib{-m}) \frac{\ib{P}^2}{\ib{K}^2} \bigg)
+ \mathcal{O}(\epsilon^{3/2}, \mathcal{F}^{3/2})
\label{eq:fft}
\end{multline}
(cf. \eq\nr{eq:bfklft}).
A diagrammatic interpretation of this as a typical BFKL ladder is given in \fig\ref{fig:ladder}. Equation~\nr{eq:momspladder} is the main result of this section. It shows that the IT Langevin equation formalism reduces, in the dilute limit, to a conventional correlation between two particles produced from the same  BFKL ladder.

The initial condition for the evolution can be read off from the definition \nr{eq:defFcal}:
\begin{equation}
\mathcal{F}^0_{\ib{x,\bar{x},u,\bar{u}}} 
= (\nc^2 -1) \delta_\ib{xu} \delta_\ib{\bar{x}\bar{u}},
\end{equation}
or in momentum space,
\begin{equation}
\mathcal{F}^0(\ib{P,\bar{P},m,\bar{m}})
= (\nc^2 -1) \delta^{(2)}(\ib{P}+\ib{m}) \delta^{(2)}(\ib{\bar{P}}+\ib{\bar{m}}).
\end{equation}
Using this in the general expression \nr{eq:momspladder} reduces the equal rapidity cross section to a $k_T$-factorized expression for the two-particle production cross section:
\begin{equation}
\left. 
\frac{d\sigma_{qg}}{dY d^2\ib{p} \, dY_A d^2\ib{k}_A}
\right|_{Y=Y_A}
= 
- \frac{\as}{(2 \pi)^2} \frac{(\ib{p}+\ib{k}_A)^2}{\ib{p}^2\ib{k}_A^2}
\varphi^0(\ib{p}+\ib{k}_A).
\label{eq:equalrap}
\end{equation}
This expression already has the structure of \fig\ref{fig:ladder}, with an unintegrated gluon distribution $\ib{q}^2 \varphi(\ib{q})$ for the gluon taken from the target, and propagators $1/\ib{k}_A^2$ and $1/\ib{p}^2$ corresponding to the produced particles. For this case of no evolution in rapidity, the Lipatov vertices for producing the gluon and the initial condition for the BFKL Green's function combine to produce a simple factor $\delta^{(2)}(\ib{q}-\ib{k}_A-\ib{p})$, which naturally means that in the absence of gluon emission in a ladder, the transverse momentum of the two final-state particles must match that coming from the target.

\section{Conclusions}

In conclusion, we have attempted to clarify here the Langevin formulation~\cite{Iancu:2013uva} of two-particle correlations in JIMWLK evolution, in the case of a dilute probe scattering off a dense color field target. Our first important result is the observation that, although the JIMWLK evolution for the Wilson lines is nonlinear, the evolution of the Lie derivatives encoding the correlation between the two rapidities, is in fact not. It can be expressed instead, with a suitable transformation, as a linear equation that is independent of the Wilson lines. This observation seems to confirm the result obtained earlier (in a rather different language) in \cite{JalilianMarian:2004da}. This conclusion points toward a much simpler way to calculate the two-particle cross section than a straightforward numerical solution of the bilocal Langevin equations in \cite{Iancu:2013uva}. Exploring the full phenomenological consequences of this observation is beyond the scope of this paper, but it would be valuable to pursue this in future work.

We have also calculated explicitly the dilute limit of the Langevin formulation, where the decorrelations in azimuthal angle between the two particles are given by a BFKL Green's function between the two rapidities. The physical picture here is that of color charges, or rather Reggeized gluon fields $\lambda$, that change with rapidity in a stochastic process (see e.g. \eqs\nr{langevin-dilute} and \nr{eq:rlambda}). The power counting for calculating two-particle correlations in this limit is, however, tricky. This is due to the fact that the two-particle cross section depends on Reggeized gluon fields from the expansion of both the adjoint Wilson lines and the Lie derivatives in the production Hamiltonian. In order to see this connection, it is easier to continue a bit further with the more formal definition in terms of the functional derivatives. The essential feature is that JIMWLK evolution as a function of the quark rapidity $Y$ ``commutes'' with the production Hamiltonian and only operates on the dipole operator at $Y$. The evolution of the double inclusive cross section with $Y$ is therefore determined by the evolution of the single inclusive cross section, but with a more complicated initial condition. This feature  enabled us to show that in the linearized limit, the result is in fact what one would expect from BFKL dynamics.

\begin{acknowledgments}
We are grateful to R. Boussarie, M. Lublinsky, E. Iancu and D. Triantafyllopoulos for discussions.
 T.~L. has been supported by the Academy of Finland, projects No. 267321 and No. 303756. A.~R. is supported by the National Research Foundation of South Africa. This work has been supported by the European Research Council, grant ERC-2015-CoG-681707.
\end{acknowledgments}

\bibliography{spires}
\bibliographystyle{JHEP-2modlong}

\end{document}